\documentclass[physrev,reprint,superscriptaddress,bibnotes]{revtex4-2}  

\usepackage[T1]{fontenc}
\usepackage[utf8]{inputenc} 
\usepackage[australian]{babel}
\usepackage{booktabs}
\usepackage[a4paper,centering,hmargin=1.75cm,vmargin=2cm]{geometry} 
\usepackage{amsmath,amssymb,graphicx,bm,microtype} 
\usepackage[colorlinks,allcolors=blue!50!black]{hyperref}
\usepackage[all]{hypcap}
\usepackage{cleveref,braket,siunitx}
\usepackage[version=4]{mhchem}
\usepackage{cleveref}
\usepackage{tikz}
\usepackage{lipsum}

\crefname{figure}{Fig.}{Figs.}
\Crefname{figure}{Figure}{Figures}
\crefname{equation}{Eq.}{Eqs.}
\Crefname{equation}{Equation}{Equations}
\crefname{section}{Sec.}{Secs.}
\Crefname{section}{Section}{Sections}

\makeatletter
\AddToHook{cmd/appendix/before}{\def\cref@section@alias{appendix}\def\cref@subsection@alias{appendix}}
\makeatother

\begin{document}
\title{Universal Quantum Control of High-Dimensional Atomic Spins via Stimulated Raman Transitions}

\author{Hon-Kwan~Chan}
\email{honkwan.chan@sydney.edu.au}
\affiliation{School of Physics, University of Sydney, NSW 2006, Australia}
\affiliation{ARC Centre of Excellence for Quantum Computer Performance \& Integration (QPI)}

\author{Vassili G.~Matsos}
\affiliation{School of Physics, University of Sydney, NSW 2006, Australia}
\affiliation{ARC Centre of Excellence for Quantum Computer Performance \& Integration (QPI)}

\author{Prachi Nagpal}
\affiliation{School of Physics, University of Sydney, NSW 2006, Australia}
\affiliation{ARC Centre of Excellence for Quantum Computer Performance \& Integration (QPI)}

\author{Teerawat Chalermpusitarak}
\affiliation{School of Physics, University of Sydney, NSW 2006, Australia}
\affiliation{ARC Centre of Excellence for Quantum Computer Performance \& Integration (QPI)}
\affiliation{Sydney Nano Institute, University of Sydney, NSW 2006, Australia}

\author{Cameron McGarry}
\affiliation{School of Physics, University of Sydney, NSW 2006, Australia}
\affiliation{ARC Centre of Excellence for Quantum Computer Performance \& Integration (QPI)}
\affiliation{Sydney Nano Institute, University of Sydney, NSW 2006, Australia}

\author{Frank~Scuccimarra}
\affiliation{School of Physics, University of Sydney, NSW 2006, Australia}
\affiliation{ARC Centre of Excellence for Quantum Computer Performance \& Integration (QPI)}

\author{Christophe H.~Valahu}
\affiliation{School of Physics, University of Sydney, NSW 2006, Australia}
\affiliation{ARC Centre of Excellence for Quantum Computer Performance \& Integration (QPI)}
\affiliation{Sydney Nano Institute, University of Sydney, NSW 2006, Australia}

\author{Ting Rei~Tan}
\email{tingrei.tan@sydney.edu.au}
\affiliation{School of Physics, University of Sydney, NSW 2006, Australia}
\affiliation{ARC Centre of Excellence for Quantum Computer Performance \& Integration (QPI)}
\affiliation{Sydney Nano Institute, University of Sydney, NSW 2006, Australia}

\date{\today}
\begin{abstract}
Zeeman sublevels of atomic angular momentum (AM) manifolds are promising qudits for quantum simulation and error correction, but controlling them becomes increasingly demanding as the manifold dimension grows. We show that an optimally controlled two-photon stimulated Raman transition provides universal control over AM manifolds of arbitrary dimension, provided it has a total electronic AM spin of $J\geq1$. Our scheme exploits the nearly degenerate energy splitting between adjacent Zeeman sublevels. We numerically demonstrate arbitrary state preparation and unitary evolution in both integer- and half-integer-spin manifolds, with errors limited by photon scattering. The scheme is applicable to both trapped-ion and neutral-atom platforms. For trapped ions, the spin-motion coupling provided by the Raman transition further enables qudit entangling operations, and we present a protocol for constructing native entangling gates on logical qubits encoded in the manifold. Our approach offers a scalable route to control high-dimensional qudits.
\end{abstract}

\maketitle

Quantum many-level systems provide access to Hilbert spaces larger than those available from two-level systems, with opportunities ranging from simulating quantum dynamics~\cite{Meth2025,John2026,champion2025lmg} to quantum information processing~\cite{Gottesman2001,Wang2020}. In atomic systems, the magnetic Zeeman sublevels within a single angular momentum (AM) manifold provide a natural realization of such systems.
The enlarged state space can encode a qudit, providing a hardware-efficient route to quantum error correction~\cite{Pirandola2008,Chiesa2020,Albert2020,Gross2021,Gross2024,Lim2023,Lim2026,Omanakuttan2024,Jain2024,Aydin2025,Furey2024,DeBry2026,Li2026} and simulations of gauge theories~\cite{Meth2025,John2026}.

Universal control of qudits requires implementing arbitrary $SU(d)$ operations within the encoded manifold. 
Existing approaches can be understood as supplementing $SU(2)$ rotations with additional controls that break the spectral degeneracy of the manifold. 
This can be achieved directly by spectrally resolving individual transitions and applying sequences of Givens rotations
~\cite{Brennen2005,OLeary2006,Neeley2009,Blok2021,Barnes2022,Krondorfer2026,Ahmed2025,Krondorfer2025a,Krondorfer2025b,Heizenreder2026,Castro2022,Ringbauer2022,Hrmo2023,Low2025,Low2026,Burshtein2026,Gregory2026,Shi2026,Champion2025,Yu2025,Lim2025,FernandezDeFuentes2024,Wang2025TransmonQudit}, or by complementing an optimally controlled collective drive of the Zeeman sublevels with an additional symmetry-breaking field~\cite{Merkel2008,Smith2013,Merkel2009,Anderson2015,Mischuck2012,Chaudhury2007,Omanakuttan2021}.

\begin{figure}[t]
    \includegraphics[width=1\linewidth]{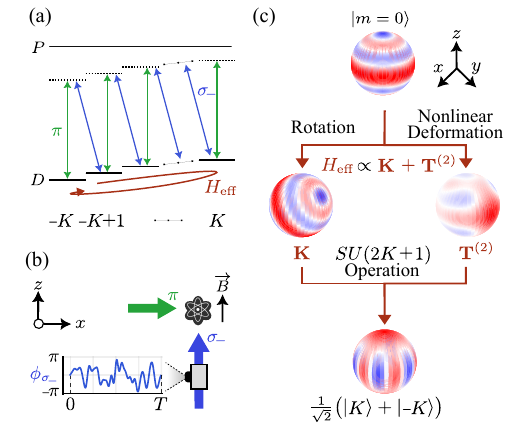}
    \caption{\label{RamanProtocolIllustration} Universal control of an AM manifold with an optimally phase-modulated two-photon stimulated Raman transition. \textbf{(a)} Nearest-neighbor magnetic Zeeman sublevels in an AM manifold of total spin $K$ within a long-lived metastable $D$ level are coherently coupled through the two-photon electric-dipole stimulated Raman transition and described by an effective Hamiltonian $H_\text{eff}$. \textbf{(b)} Universal control of an AM manifold is achieved using optimal phase modulation on one of the light fields enacting the two-photon Raman transition. \textbf{(c)} The effective Hamiltonian $H_\text{eff}$ describing the two-photon Raman transition contains two necessary AM operations for universal control. For instance, preparing the $\ket{m=0}$ state into a spin-cat state is an $SU(2K+1)$ operation. The effective Hamiltonian consists of an AM rotation described by $\mathbf{K}$ and a nonlinear AM operation described by the rank-$2$ STO $\mathbf{T}^{(2)}$. By phase modulating the effective Hamiltonian, the combination of both AM operations has arbitrary $SU(2K+1)$ control.}
\end{figure}

Here, we add a complementary control toolkit based on laser-driven two-photon stimulated Raman transitions that intrinsically provides universal controllability without necessarily breaking the spectral degeneracy of individual qudit transitions.
Our approach leverages the 
asymmetric, Zeeman-state-dependent Rabi frequencies across an AM manifold and uses optimal collective drive for universal qudit control to generate arbitrary $SU(d)$ operations. 
The minimum control-resource requirement remains constant with qudit dimension: as few
as two optical fields can provide universal control of an arbitrarily large AM manifold, provided their Raman couplings connect all Zeeman sublevels. Furthermore, the Raman interaction keeps the population within the computational Hilbert space and provides flexibility to suppress AC Stark shifts or enable their calibration. 
The approach is applicable to both trapped-ion and neutral-atom architectures. Moreover, in trapped ions, the scheme naturally provides spin-motion coupling for implementing qudit entangling operations.

We first establish the universal controllability of Zeeman sublevels driven with the two-photon Raman transition. We then demonstrate the protocol on representative integer- and half-integer-spin systems, achieving arbitrary state preparation and unitary transformations with infidelities limited by photon scattering. Finally, as an application of our protocol, 
we construct a universal gate set for logical qubits encoded using the absorption-emission (AE) scheme~\cite{Jain2024,Aydin2025}.

\section{Theory}
We discuss the requirement on physically accessible operators for universal control. We consider an arbitrary AM $K$, where $K$ may denote either the hyperfine AM $F$ or the total electronic AM $J$. The AM manifold is comprised of $d=2K+1$ Zeeman sublevels, $\ket{K,m}$, where $m = -K, -K+1, ..., K$. We show mathematically that the two-photon electric-dipole Raman transition fulfills the requirement for universal control for both hyperfine and total electronic spin manifolds, provided that $J$  $\geq1$.

Universal control demands a set of physical operators that generate the $su(d)$ Lie algebra. This requires that linear combinations and commutators of the physical operators have support on spherical tensor operators $T^{(k)}_q(K)$ of rank $k=1$ to $2K$, which together with the identity ($k=0$) form a complete operator basis for the $d$-dimensional Hilbert space~\cite{Merkel2008}. 

The general rank-$k$ spherical tensor operator (STO) is
\begin{align}
        T^{(k)}_q(K) &= \sqrt\frac{2k{+}1}{2K{+}1}\sum_m C^{K,m{+}q}_{k,q;K,m} |K,m{+}q\rangle\langle{K,m}|,
\end{align}
with $C^{K,m{+}q}_{k,q;K,m}=\langle K,m{+}q|k,q;K,m\rangle$ denoting the Clebsch-Gordan (CG) coupling coefficient, and $q\in[-k,k]$ labels the component of the STO of rank $k$. 

The $x,y$ angular momentum operators 
\begin{align}
    K_x&\propto\sum_mC^{K,m+1}_{1,1;K,m}\ket{K,m{+}1}\bra{K,m} +\text{h.c.}, \notag \\
    K_y&\propto\sum_m-iC^{K,m+1}_{1,1;K,m}\ket{K,m{+}1}\bra{K,m} +\text{h.c.}.
\end{align}
are Hermitian linear combinations of rank-1 STOs. The linear combinations and commutations of $K_{x}$ and $K_y$ remain rank-1 and therefore generate only $su(2)$ algebra. Extension to $su(d)$ algebra requires an additional operator $O$ that has a non-zero overlap with a rank-2 STO, i.e.,
\begin{align}
    \mathrm{Tr}\big(O^\dagger \, T_q^{(2)}(K)\big) &\neq 0.\label{UniversalControlCriteria}
\end{align}
The set $\{K_x, K_y, O\}$ generates, through linear combinations and commutators, all STOs of rank $k\geq1$ and hence the full $su(d)$ algebra required for universal control~\cite{Merkel2008}.

We show analytically that the two-photon electric-dipole Raman transitions comprise up to rank-2 STOs which physically realize the control operators $\{K_x,K_y,O\}$ required for control universality (Fig.~\ref{RamanProtocolIllustration}). We assume that an AM manifold is subject to a sufficiently weak external magnetic field such that the quadratic Zeeman shifts are negligible. The remaining linear Zeeman shift then produces a uniform energy spacing $\omega_0$ between neighboring Zeeman sublevels.

We consider Raman light fields interacting with a lower and an upper excited AM manifold of spin $K$ and $K'$. The light-atom interaction can be described by the Hamiltonian $H = H_0+V(t)$
where the bare Hamiltonian, $H_0$, and the laser-induced coupling, $V(t)$, are 
\begin{eqnarray}
    H_0&=&\sum_mm\omega_0\ket{K,m}\bra{K,m} \notag \\ &&+\sum_{m'}\big(m'\omega'_0+\omega_{e}\big)\ket{K',m'}\bra{K',m'},\\
    V(t)&=&\sum_i\mathbf{d}_{K',K}^{+} {\cdot} E_i\hat{\varepsilon}_{i}e^{-i\omega_i t}+\mathbf{d}_{K',K}^{-} {\cdot} E_i\hat{\varepsilon}_{i}^*e^{i\omega_i t}.\label{lightatomcouplingbare}
\end{eqnarray}
In these expressions, $m'$ denotes the Zeeman sublevels for the $K'$ excited AM manifold, $\omega_0'$ and $\omega_e$ are the Zeeman splitting in $K'$ and the energy difference between the $K'$ and $K$ levels, and $E_i$, $\hat{\varepsilon}_i$, $\omega_i$ denote the electric field strength, polarization, and frequency of the $i$th Raman light field.
We have made the rotating-wave approximation in Eq.~\ref{lightatomcouplingbare}, where $\mathbf{d}_{K',K}^{+}=\sum_{m,m'}\bra{K',m'}\mathbf{d}\ket{K,m}\ket{K',m'}\bra{K,m}$ is the part of the electric-dipole operator that raises the population from $K$ to $K'$, and $\mathbf{d}_{K',K}^{-}=\left(\mathbf{d}_{K',K}^{+}\right)^{\dagger}=\sum_{m,m'}\bra{K,m}\mathbf{d}\ket{K',m'}\ket{K,m}\bra{K',m'}$ is the part that lowers it from $K'$ to $K$. 

We perform a unitary transformation $U=e^{-iH_0 t}$ into the interaction picture with respect to $H_0$ and adiabatically eliminate the excited states $\ket{K',m'}$ in the far detuned regime to obtain the effective light-atom Hamiltonian for the two-photon stimulated Raman transition~\cite{James2007,Steck,Deutsch2010,LeKien2013} 
\begin{align}
H_{\rm eff} ={}& \sum_{i,j,m,m''}
  \frac{E_iE_j}{4\bar{\Delta}_{ij}}\,
  e^{-i[\omega_{ij}-(m''-m)\omega_0]t}\,
  \ket{K,m''}\bra{K,m} \nonumber\\
  &\times \bra{K,m''}(\hat{\varepsilon}_j^{*}\cdot\mathbf{d}^{-}_{K',K})
  (\mathbf{d}^{+}_{K',K}\cdot\hat{\varepsilon}_i)\ket{K,m},
\label{effectiveHamitlonianCompact}
\end{align}
where $\omega_{ij} = \omega_i - \omega_j$,
and $\bar{\Delta}_{ij}$ is the harmonic mean of the detunings of the two light fields,
given by $2/\bar{\Delta}_{ij} = 1/\Delta_i + 1/\Delta_j$, with
$\Delta_{i(j)} = \omega_{i(j)} - \omega_e$. We have dropped the dependence on the Zeeman splittings in $\bar{\Delta}$ by assuming that the Zeeman splittings are much smaller than the detunings, i.e., $\omega_0,\omega_0' \ll \Delta_{i(j)}$.  
Equation~\ref{effectiveHamitlonianCompact} is general and applicable to configurations involving more than two light fields to determine the effective Hamiltonian of the two-photon Raman transition.

We now consider a pair of light fields labeled by $\mu$ and $\nu$, where the effective Hamiltonian in Eq.~\ref{effectiveHamitlonianCompact} can be written as
\begin{align}
    H_{\text{eff}}&= \sum_{a=\{\mu,\nu\}}\sum_m\Omega_m^{a,a}\ket{K,m}\bra{K,m} + \notag \\
        &\quad \sum_m \Omega_m^{\nu,\mu}e^{-i(\omega_{\mu\nu}-q\omega_0)t}\ket{K,m{+}q}\bra{K,m} + \text{h.c.} \big)~\label{effHamiltonian_munu},
\end{align}
where the first term describes the AC Stark shift induced by the individual light field, while the second describes a state-changing two-photon Raman transition with $\Delta m = q$. 

Each coupling term in the effective Hamiltonian given in Eq.~\ref{effHamiltonian_munu} can be compactly represented by STOs. To illustrate this, consider the case where the light fields have pure polarizations in the spherical basis, i.e., $\hat{\varepsilon}_\mu,\hat{\varepsilon}_\nu\in\{-1,0,+1\}$, corresponding to left circular ($\hat{\sigma}_-$), linear ($\hat{\pi}$), and right circular ($\hat{\sigma}_+$) polarizations, respectively. The corresponding coupling strengths are given by
\begin{align}
    \Omega_m^{\nu,\mu}&=\frac{E_\mu E_\nu}{4\bar{\Delta}_{\mu,\nu}}\bra{K,m{+}q}(\hat{\varepsilon}_\nu^{*}\cdot\mathbf{d}^{-}_{K',K})
  (\mathbf{d}^{+}_{K',K}\cdot\hat{\varepsilon}_\mu)\ket{K,m} \notag \\
    &=(-1)^\nu\frac{E_\mu E_\nu}{4\bar{\Delta}_{\mu,\nu}}\sum_{m'}\langle K,m{+}q|d_{-\nu}|K',m'\rangle \notag \\ &\quad \times \langle K',m'|d_{\mu}|K,m \rangle,\label{Two-PhotonRabiFrequencyEquation}
\end{align}
where terms with $\mu=\nu$ give the AC Stark shift due to individual light fields. 
By considering the properties of STO products and the Wigner-Eckart theorem, we can write the summation of the two-photon coupling $\Omega_m^{\nu,\mu}\ket{K,m{+}q}\bra{K,m}$ across the $K$ manifold in terms of STOs up to rank-2 $($see appendix \ref{RabiFrequencyDecompositionIntoSphericalTensors}$)$  
\begin{align}
\sum_{m}\Omega_m^{\nu,\mu}\ket{K,m{+}q}\bra{K,m} =
    (-1)^\nu\frac{E_\nu E_\mu}{4\bar{\Delta}_{\mu,\nu}} \notag \\ \times \sum_{k=0}^2C^{k,q}_{1,-\nu,1,\mu}A^{(k)}(K) T^{(k)}_q(K),\label{CouplingStrengthFinalDecomposition}
\end{align}
where $A^{(k)}(K)$ is a coefficient that determines the strength of the rank-$k$ component of the effective two-photon coupling within the $K$ manifold. For systems with hyperfine AM $(K\equiv F)$, this coefficient is
\begin{align}
    A^{(k)}(F)&=(-1)^{-F-2J-J'-I+k}(2F+1)(2J+1)\notag \\ &\times\left\{
\begin{matrix}
1 & 1 & k \\
J & J & J'
\end{matrix} \right\}\left\{
\begin{matrix}
J & J & k \\
F & F & I
\end{matrix} \right\}|\langle J||\mathbf{d}||J'\rangle|^2,\label{Coeff_hyperfine}
\end{align}
where $\left\{\begin{matrix}
* & * & * \\
* & *& *
\end{matrix}\right\}$ denotes the Wigner-$6j$ symbol. We further assume that the detuning $\bar{\Delta}$ is much larger than the hyperfine splittings of the excited level $J'$. For systems without nuclear spin $(K\equiv J)$,
\begin{align}
    A^{(k)}(J)&=(-1)^{2k+J+J'}(2J+1) \notag \\ 
    &\times\left\{
\begin{matrix}
1 & 1 & k \\
J & J & J'
\end{matrix} \right\}|\langle J||\mathbf{d}||J'\rangle|^2 .\label{Coeff_fine}
\end{align}

The physical interpretation of Eq.~\ref{CouplingStrengthFinalDecomposition} becomes apparent by grouping terms according to $q=\mu-\nu$, which corresponds to the change in the magnetic quantum number. The $q=0$ term describes the AC Stark shift, where the rank-$0$, $1$, and $2$ STOs describe scalar, vector, and tensor shifts, respectively. The $q=\pm2$ terms drive $|\Delta m|=2$ transitions and are purely rank-$2$. 
Of particular interest is the $q=\pm1$ term. This $|\Delta m|=1$ transition contains both a rank-$1$ component, which forms the angular momentum operators $K_{x,y}$, and can also contain a non-zero rank-$2$ component. Together, the rank-1 and rank-2 components provide the required operators $\{K_x,K_y,O\}$ for control universality. Thus, a pair of Raman light fields driving $|\Delta m|=1$ transitions can generate the
complete control set, provided the Raman couplings connect all Zeeman sublevels of the manifold.

Within the far-detuned regime considered here, universal controllability via the Raman transition requires the AM manifold to satisfy $J\geq1$. 
This is because the Wigner-6$j$ symbol (in Eq.~\ref{Coeff_hyperfine} and Eq.~\ref{Coeff_fine}) imposes triangle constraints such that $\left\{\begin{matrix} 1 & 1 & k \\ J & J & J' \end{matrix}\right\}$ vanishes when $J<1$ for $k=2$. 
For example, the $S_{1/2}$ ground levels of alkali-like ions do not meet this condition, whereas AM manifolds in long-lived metastable $D$ levels typically have $J\geq1$ and are therefore universally controllable via Raman coupling.
We note that this restriction assumes the Raman detuning $\bar{\Delta}$ is much larger than the excited-state hyperfine splitting. 
When the detuning is comparable to this splitting, transitions with $J<1$ can also retain rank-2 STO contributions~\cite{Krondorfer2026,Ahmed2025,Yang2024,Kusano2026SpinCat}.

\section{Examples}\label{examples section}

\begin{figure}[t]
    \includegraphics[width=1\linewidth]{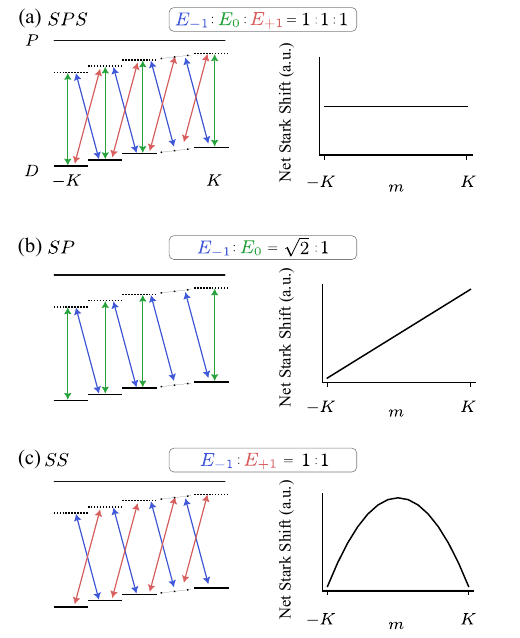}
    \caption{\label{QuditRamanConfig}Raman light field configurations for universal control of an AM manifold and the associated net Stark shift. \textbf{(a)} $(SPS)$ $\sigma_-,~\pi,~\sigma_+$ polarized fields of equal strength yield a constant (scalar) Stark shift. \textbf{(b)} $(SP)$ $\sigma_-,~\pi$ polarized fields with strength ratio $E_{-1}:E_0=\sqrt{2}:1$ yield a linear (vector) Stark shift. \textbf{(c)} $(SS)$ $\sigma_-,~\sigma_+$ polarized fields of equal strength yield a quadratic (tensor) Stark shift.}
    \label{CouplingandSSFigure}
\end{figure}

Having established that two-photon stimulated Raman transitions can provide the rank-1 and 2 operators required for control universality, 
we consider three Raman light field configurations labeled as $SPS$, $SP$, and $SS$. Together, the three configurations constitute all possible two-photon Raman couplings on the AM manifold (Fig.~\ref{QuditRamanConfig}). The most general, $SPS$, uses three fields with $\sigma_-$, $\pi$, and $\sigma_+$ polarizations and drives both $|\Delta m|=1$ and $|\Delta m|=2$ transitions. The $SP$ configuration uses ($\sigma_-$, $\pi$) fields and drives only $|\Delta m|=1$ transitions, while $SS$ uses ($\sigma_-$, $\sigma_+$) fields and drives only $|\Delta m|=2$ transitions.
The general effective Hamiltonian for the $SPS$ configuration is given by 

\begin{widetext}
    \begin{align}
         H^{SPS}_{\text{eff}}(K) &= \Big(\sum_{k=1}^2 \big(\frac{(-1)^{-1}E_0E_{-1}}{4\bar{\Delta}_{0,-1}}C^{k,1}_{1,1,1,0}A^{(k)}T^{(k)}_1e^{-i\delta_{0,{-}1,1} t + i\phi_{0,-1}} + \frac{(-1)^{0}E_{+1}E_{0}}{4\bar{\Delta}_{1,0}}C^{k,1}_{1,0,1,1}A^{(k)}T^{(k)}_1e^{-i\delta_{{+}1,0,1} t + i\phi_{+1,0}} \big) \notag \\ &+ \frac{(-1)^{-1}E_{+1}E_{-1}}{4\bar{\Delta}_{1,-1}}C^{2,2}_{1,1,1,1}A^{(2)}T^{(2)}_2e^{-i\delta_{{+}1,{-}1,2}t + i\phi_{+1,-1}} + \text{h.c.} \Big) +  \sum_{i=-1}^1\sum_{k=0}^2 \frac{(-1)^{i}E_i^2}{4\Delta_i}C^{k,0}_{1,-i,1,i}A^{(k)}T^{(k)}_0. \label{EffectiveHamiltonian}
    \end{align}
\end{widetext}
Here, $E_i$ denotes the electric field strength of polarization $i$, $\delta_{i,j,q}=\omega_{ij}-q\omega_0$ and $\phi_{i,j}=\phi_i-\phi_j$ is the phase difference between the purely polarized fields $i$ and $j$. The first and second terms drive $|\Delta m|=1$ transitions via the ($\sigma_-$, $\pi$) and ($\pi$, $\sigma_+$) polarization pairs, the third term drives $|\Delta m|=2$ transitions via ($\sigma_-$, $\sigma_+$), and the last term is the net AC Stark shift from all three polarizations. The $SP$ and $SS$ Hamiltonians follow by setting the absent polarization's field amplitude to $E_{+1}=0$ and $E_{0}=0$, respectively.

The state-changing Raman transition terms in $H^{SPS}_{\text{eff}}(K)$ carry the relative phases $\phi_{i,j}=\phi_i-\phi_j$ of the participating light fields. Since only phase differences enter, these are not all independent: the phase of the $|\Delta m|=2$ term is fixed by the two $|\Delta m|=1$ terms, $\phi_{+1,-1}=\phi_{+1,0}+\phi_{0,-1}$. The $SPS$ configuration therefore provides two independent controls, $\phi_{+1,0}$ and $\phi_{0,-1}$, whereas the $SP$ and $SS$ configurations each provide one. We implement optimal control by modulating these relative phases while driving the transitions resonantly, i.e.\ $\delta_{i,j,q}=0$
for each pair of polarizations $i$ and $j$ (see appendix~\ref{PhaseModulationOptimalControl}).

The relative electric-field amplitudes ($E_i$) associated with the $i$th polarization component provide an additional control over the net AC Stark shift generated by the Raman fields. By choosing $E_i$ appropriately, we can selectively enhance or cancel the scalar, vector, and tensor contributions to the Stark shift. Fig.~\ref{CouplingandSSFigure} illustrates this for the three Raman configurations considered here. For the $SPS$ configuration, equal field amplitudes produce a purely scalar Stark shift that is independent of the magnetic quantum number, $m$. For the $SP$ configuration, setting $E_{-1}:E_0=\sqrt{2}:1$ cancels the tensor contribution, leaving a vector Stark shift that varies linearly with $m$. For the $SS$ configuration, setting $E_{-1}:E_{+1}=1:1$ cancels the vector contribution and produces a tensor Stark shift with quadratic dependence on $m$. These well-defined Stark shift profiles across the manifold also provide a convenient experimental means to calibrate the relative intensities and detunings in $H_{\text{eff}}(K)$~\cite{Gregory2026}. 

We implement universal control on two representative manifolds: the metastable $^3D_1$ level of $^{176}\text{Lu}^+$ with integer AM $F=8$, and the $^2D_{5/2}$ level of $^{40}\text{Ca}^+$ with half-integer AM $J=5/2$. The $^{176}\text{Lu}^+$ manifold has a linear Zeeman splitting of $\omega_0/(2\pi B)\approx 87~\text{kHz/G}$, and Raman transitions are driven via far-off-resonant coupling to the $^3P_0$ and $^3P_1$ excited states, with detunings $\Delta/2\pi \approx 1.4~\text{THz}$ and $-35.7~\text{THz}$, respectively. The $^{40}\text{Ca}^+$ manifold has a linear Zeeman splitting of $\omega_0/(2\pi B)\approx 1.68~\text{MHz/G}$, and Raman transitions are mediated through the $^2P_{3/2}$ excited states with $\Delta/2\pi \approx -43.7~\text{THz}$. In both systems the Zeeman splittings are negligible compared to the Raman detunings, so we set $\bar{\Delta}_{i,j}=\Delta_i=\Delta$ for all polarizations $i$ and $j$.

\subsection{Arbitrary State Preparation and Unitary Evolution}\label{Arbitrary State Preparation and Unitary Evolution}
We begin by generating several nontrivial spin states in the $^{176}\text{Lu}^+$ $F=8$ and $^{40}\text{Ca}^+$ $J=5/2$ AM manifolds that are relevant for hardware-efficient logical encoding or single-atom quantum error correction. We take the initial states to be $\psi_\text{ini}=\ket{8,0}$ and $\ket{5/2,-1/2}$ for $^{176}\text{Lu}^+$ and $^{40}\text{Ca}^+$ respectively. 

For each Raman configuration, the dimension $d$ of the preparable states is set by the Zeeman sublevels that can be reached through the available Raman coupling from the given initial state. The $SPS$ configuration connects every sublevel of the AM manifold, giving access to the full space with $d=17$ for $^{176}\text{Lu}^+$ and $d=6$ for $^{40}\text{Ca}^+$. The $SP$ configuration likewise reaches the full manifold of $^{176}\text{Lu}^+$ ($d=17$), but accesses only $d=5$ sublevels of $^{40}\text{Ca}^+$, namely $m\in\{-3/2,\dots,5/2\}$, because the $^2P_{3/2}$ manifold has no excited sublevel that can mediate a Raman coupling to $m=-5/2$. In the $SS$ configuration, the coupling allows only $|\Delta m|=2$ transitions, so starting from $\ket{8,0}$ ($\ket{5/2,-1/2}$) it reaches only the sublevels whose $m$ differs from the initial value by an even integer: $m\in\{-8,-6,\dots,8\}$ for $^{176}\text{Lu}^+$ and $m\in\{-5/2,-1/2,3/2\}$ for $^{40}\text{Ca}^+$. These form subspaces of dimension $d=9$ and $d=3$, respectively. Each AM subspace is closed under the $SS$ couplings, within which the $SS$ configuration provides universal control (see appendix~\ref{Control Universality}). 

We prepare three classes of states. First, for both ion species we prepare the spin-cat state~\cite{DeBry2026,Omanakuttan2024}, an equal superposition of the extremal $m$ states,
\begin{align}
\ket{\text{cat}}_{5/2} &= \tfrac{1}{\sqrt{2}}\left(\ket{\tfrac{5}{2},\tfrac{5}{2}}+\ket{\tfrac{5}{2},-\tfrac{5}{2}}\right), \notag \\
\ket{\text{cat}}_{8} &= \tfrac{1}{\sqrt{2}}\left(\ket{8,8}+\ket{8,-8}\right). \label{eq:cat}
\end{align}

Second, for $^{40}\text{Ca}^+$ we prepare the spin-$5/2$ binary octahedral (BO) irreducible-representation codewords~\cite{Gross2021},
\begin{align}
\ket{\bar0}_\text{BO} &= \sqrt{\tfrac{1}{6}}\,\ket{\tfrac{5}{2},\tfrac{5}{2}} - \sqrt{\tfrac{5}{6}}\,\ket{\tfrac{5}{2},-\tfrac{3}{2}}, \notag \\
\ket{\bar1}_\text{BO} &= -\sqrt{\tfrac{5}{6}}\,\ket{\tfrac{5}{2},\tfrac{3}{2}} + \sqrt{\tfrac{1}{6}}\,\ket{\tfrac{5}{2},-\tfrac{5}{2}}. \label{eq:BO}
\end{align}

Third, for $^{176}\text{Lu}^+$ we prepare the absorption-emission (AE) codewords~\cite{Jain2024},
\begin{align}
\ket{\bar0}_\text{AE} &= \sqrt{\tfrac{3}{4}}\,\ket{8,-2} + \sqrt{\tfrac{1}{4}}\,\ket{8,6}, \notag \\
\ket{\bar1}_\text{AE} &= \sqrt{\tfrac{3}{4}}\,\ket{8,2} + \sqrt{\tfrac{1}{4}}\,\ket{8,-6}. \label{AECodeword}
\end{align}

For these target states, we achieve state-preparation infidelities, $\mathcal{I}_{\text{s}} = 1-|\langle\psi_\text{tar}|\psi_\text{prep}\rangle|^2$, below $10^{-6}$. A more fundamental limit arises from spontaneous photon scattering induced by the off-resonant Raman light fields, which constitutes an intrinsic error source for coherent operations driven by stimulated Raman transitions~\cite{Ozeri2007,Moore2023}. We estimate this error by integrating the excited-state population, weighted by the decay rate, over the pulse duration, obtaining values on the order of $10^{-3}$ and $10^{-5}$ for $^{176}\text{Lu}^+$ and $^{40}\text{Ca}^+$, respectively, which exceed the coherent state-preparation infidelity. Details on the photon-scattering error estimate and state-preparation results are given in Appendix~\ref{PhotonScatterEstimate}. 

\begin{figure}[h]
    \includegraphics[width=1\linewidth]{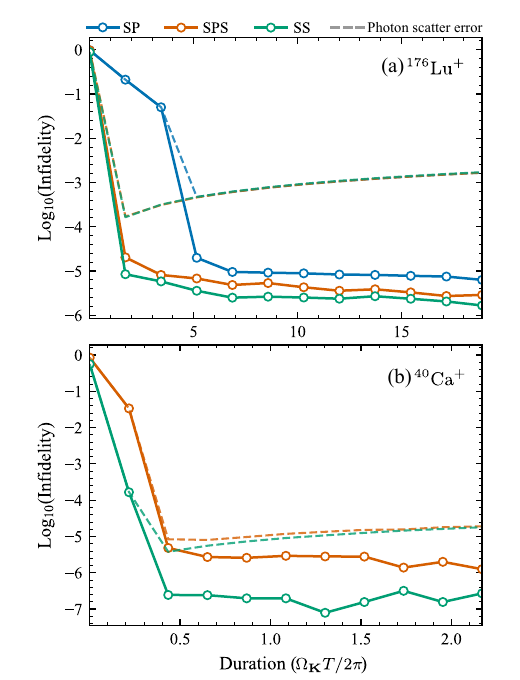}
    \caption{Average logarithmic state preparation infidelity for $^{176}\text{Lu}^+$ and $^{40}\text{Ca}^+$ versus dimensionless pulse duration $\Omega_{\mathbf{K}} T/2\pi$. Here, $\Omega_\mathbf{K}$ is the Rabi frequency of the $SU(2)$ rotation generated by the rank-1 STO component of the $SP$ configuration, which serves as a common time unit for all Raman configurations. Log infidelities are averaged over $10$ Haar-random target states. Solid lines show noiseless infidelities for each Raman configuration ($SPS$, $SP$, $SS$). Dashed lines show the estimated photon scattering error.} 
    \label{Infidelity_vs_Duruation}
\end{figure}

To further substantiate the universality of the protocol, we prepare 10 states randomly sampled according to the Haar measure within the subspace accessible to each configuration. Figure~\ref{Infidelity_vs_Duruation} shows the resulting state-preparation infidelity averaged across the 10 states as a function of the dimensionless phase-modulation duration $\Omega_\mathbf{K} T/2\pi$, for $^{176}\text{Lu}^+$ in panel~(a) and $^{40}\text{Ca}^+$ in panel~(b). Here, $\Omega_\mathbf{K}$ is the Rabi frequency of the $SU(2)$ rotation generated by
the rank-1 STO component of the $SP$ configuration, which serves as a common time unit for all Raman configurations. Each configuration is simulated with the same total optical intensity, with the relative electric-field amplitudes given in Fig.~\ref{QuditRamanConfig}. Solid curves show the infidelity from coherent evolution under the effective Hamiltonian, and dashed curves show the error due to photon scattering. Beyond a minimum pulse duration, all configurations reach coherent infidelities well below the corresponding photon-scattering error, so the achievable fidelity is set by photon scattering in every case. The configurations differ mainly in the minimum duration required and in the subspace they can access.

Beyond state preparation, we also implement Haar-random unitaries, optimizing the pulse duration together with the phase modulation. We consider two cases: the full AM manifold reached by the $SPS$ configuration and the subspace connected by the $SS$ configuration. Table~\ref{UniversalControlFidelities} reports the logarithmic infidelities for these two configurations, averaged over 10 sampled unitaries. We do not include the $SP$ configuration which also connects the full $F=8$ manifold of $^{176}\text{Lu}^+$ because optimizing a phase-modulation waveform to implement a unitary in this configuration typically takes hours to converge and fails more frequently. This is because the number of optimizable phases required in the waveform grows quadratically with the system dimension $d$ when targeting unitaries (see appendix~\ref{PhaseModulationOptimalControl}).

\begin{table*}[t]
\centering
\setlength{\tabcolsep}{8pt}
\renewcommand{\arraystretch}{0.95}
\begin{tabular}{lccccc}
\specialrule{1.2pt}{0pt}{1pt}
\specialrule{0.4pt}{1pt}{3pt}
Target & $\log_{10}$(Infidelity) & Duration ($\Omega_\mathbf{K} T/2\pi$) &  $\log_{10}$(Photon scatter error) & Raman config. & Ion\\
\specialrule{0.4pt}{2pt}{2pt}
$U_\mathrm{arb}(d=6 )$ & $-6.35$ & $1.57$ &$-4.89$ & $SPS$ &$^{40}\text{Ca}^+$ \\
$U_\mathrm{arb}$$(d=3)$ & $-7.45$ & $1.11$ &$-5.04$ & $SS$  &$^{40}\text{Ca}^+$ \\
$U_\mathrm{arb}$$(d=17)$ & $-6.69$ & $22.97$ & $-2.70$& $SPS$ &$^{176}\text{Lu}^+$\\
$U_\mathrm{arb}$$(d=9)$ & $-6.83$ & $13.93$ & $-2.90$ & $SS$  &$^{176}\text{Lu}^+$\\
$H_{\text{AE}}$ & $-6.57$ & 9.39 & $-3.08$& $SP$ &$^{176}\text{Lu}^+$  \\
$T_{\text{AE}}$  & $-7.55$ & 11.31 &$-3.00$ & $SP$  &$^{176}\text{Lu}^+$\\
\specialrule{0.4pt}{2pt}{1pt}
\specialrule{1.2pt}{1pt}{0pt}
\end{tabular}
\caption{\label{UniversalControlFidelities}Noiseless infidelities, pulse durations, and estimated photon scattering errors for unitary operations in $^{176}\text{Lu}^+$ and $^{40}\text{Ca}^+$ using different Raman configurations. For arbitrary unitary operations $U_{\text{arb}}$, infidelities and durations are averaged over ten Haar-random unitaries of dimension $d$. The single-qubit Hadamard ($H_{\text{AE}}$) and $T$ ($T_{\text{AE}}$) gates act on the AE codewords in Eq.~\ref{AECodeword}.}
\end{table*}

\subsection{Gate Set for Absorption-Emission (AE) Codeword}\label{UniversalGateSet}
We now consider implementing a universal gate set on logical qubits encoded in an AE code. AE codes are designed to protect against the errors native to AM systems, i.e., dephasing and photon absorption and emission processes such as spontaneous photon scattering induced by the Raman driving beams. By encoding the logical states as superpositions distributed across the Zeeman sublevels of a single AM manifold, these errors can be made to satisfy the Knill-Laflamme conditions for correctability~\cite{Jain2024,Aydin2025}. Thus, the number of correctable errors increases with the total spin $K$ of the host manifold; for instance, detecting up to two AE errors requires $K \geq 9/2$, while correcting up to two requires $K \geq 19/2$~\cite{Aydin2025}. Here, we use the AE codeword given in Eq.~\ref{AECodeword}, which corrects a single AE error that changes the AM by at most one unit or it can detect up to two AE errors. This code requires $K \geq 9/2$ and can therefore be encoded in the $F = 8$ manifold of $^{176}\text{Lu}^+$.

For this example, we employ the $SP$ configuration, which connects all of the Zeeman sublevels in the $F=8$ manifold with two light fields. 
We generate the logical single-qubit Hadamard ($H_\text{AE}$) and $T$ ($T_\text{AE}$) gates and obtain infidelities on the order of $10^{-3}$ limited by photon scattering errors (see Table~\ref{UniversalControlFidelities}). 

To complete the universal gate set, we now consider a logical two-qubit gate, which requires an entangling interaction between qudits encoded in the Zeeman sublevels of an AM manifold.
In trapped ions, such interactions can be mediated by a shared motional mode, where state-dependent forces drive closed trajectories in the motional phase space to generate a state-dependent geometric phase that creates entanglement. 
Similar to prior approaches to universal $SU(d)$ control, existing entangling schemes leverage interactions that break the spectral degeneracy of the encoded manifold~\cite{Ringbauer2022,Hrmo2023}. This can be achieved by directly resolving transitions so that only subsets of qudit states participate in the phase-space dynamics to create entanglement, or resolving all qudit states with multiple laser drives using a generalized M\o lmer-S\o rensen (MS) interaction~\cite{Low2020}. 

For an AM manifold with spectrally degenerate $|\Delta m| = 1$ transitions considered in this work, the generalized MS interaction is particularly natural and provides two advantages. First, because the nearest-neighbor transitions are degenerate, a single bichromatic Raman drive can address them simultaneously across the entire manifold, avoiding the need for multiple bichromatic drives for individual transitions and thereby reducing the associated spectral-control and calibration overhead. Second, this generalized MS interaction directly generates native qudit entanglement by having all qudit states within the AM manifold participate in collective spin-dependent evolutions in the motional phase space. This can reduce the number of entangling operations and the overall circuit depth required to synthesize certain target two-qudit transformations~\cite{Low2020,Ringbauer2022,Hrmo2023,Brennen2005}. 

Here, we consider generating a generalized MS interaction in the $F=8$ manifold of $^{176}\text{Lu}^+$ using the $SP$ configuration. The effective Hamiltonian including the wavevector contributions of the $\sigma_-$ and $\pi$ fields is given by
\begin{widetext}
\begin{align}
    H^{SP}_{\text{eff}}(F=8)=\sum_{k=1}^2 \big(e^{i\mathbf{k}_{0,-1}\cdot \mathbf{r}}\frac{(-1)^{-1}E_0E_{-1}}{4\bar{\Delta}_{0,-1}}C^{k,1}_{1,1,1,0}A^{(k)}T^{(k)}_1e^{-i\delta_{0,-1} t + i\phi_{0,-1}} +\text{h.c.}\big)+\sum_mm\omega_\text{Stark}\ket{8,m}\bra{8,m} ,
\end{align}
\end{widetext}
where $\mathbf{r}$ denotes the position operator for the ion, $\mathbf{k}_{0,-1}=\mathbf{k}_0-\mathbf{k}_{-1}$ is the wavevector difference for the non-copropagating fields $\sigma_-$ and $\pi$, and $\omega_\text{Stark}$ denotes the linear Stark shift splitting due to the $SP$ configuration (see Fig.~\ref{QuditRamanConfig}). By making a unitary transformation into the frame of the linear Stark shift and the harmonic oscillator, $H_m=\nu(\hat{a}^\dagger\hat{a}+\tfrac{1}{2})$, and applying the red and blue sideband interactions simultaneously at symmetric detunings from the harmonic motion frequency $\nu$, we obtain, after discarding fast-oscillating terms~\cite{Leibfried2003,Low2020}, the single-ion generalized MS Hamiltonian
\begin{align}
    H_{\text{MS}} = \eta\xi\big(a^\dagger e^{i\epsilon t} + a\,e^{-i\epsilon t}\big).
    \label{QuditSDF}
\end{align}
Here, $\mathfrak{\epsilon}=\delta_{0,-1}-\nu-\omega_\text{Stark}$ is the detuning of the Raman drive from the motional frequency and the linear Stark shift, $ \eta=|\mathbf{k}_{0,-1}|\sqrt{\hbar/2m\nu}$ is the Lamb-Dicke parameter and the Hermitian AM operator $\xi$ is given by
 \begin{align}
     \xi= i\big(\sum_{k=1}^2 \frac{(-1)^{-1}E_0E_{-1}}{4\bar{\Delta}_{0,-1}}C^{k,1}_{1,1,1,0}A^{(k)}T^{(k)}_1-\text{h.c.}\big).
 \end{align}

In the case of two ions, the single-ion AM operator $\xi$ is replaced by the collective operator $\Xi=\xi\otimes I+I\otimes \xi$, where $I$ is the identity operator. The time evolution of two ions and their collective motion under the generalized MS interaction for a duration of $t=2\pi/|\epsilon|$ constructs a qudit state-dependent geometric phase gate
\begin{align}
    U_{\text{MS}}(\Phi)=e^{i\Phi \Xi^2},
    \label{MSunitary}
\end{align}
with the entangling geometric phase $\Phi=2\pi\eta^2/\epsilon^2$.

\begin{figure}[t]
    \centering
    \includegraphics[width=\linewidth]{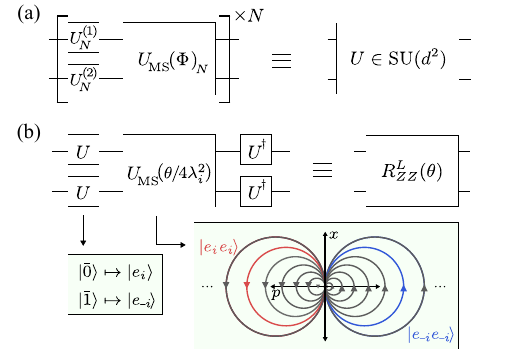}
    \caption{\label{TwoQuditCircuitDiagram}Two-qudit gate synthesis using universal control and a generalized MS interaction. \textbf{(a)} $N$ repetitions of single-qudit operations $U_N$ interleaved with generalized MS interaction $U_{\text{MS}}(\Phi)_N$ synthesize an arbitrary two-qudit unitary $U\in SU(d^2)$. \textbf{(b)} Logical $R^L_{ZZ}(\theta)$ gate. Single-qudit operations $U$ map $\ket{\bar0}\mapsto\ket{e_i}$ and $\ket{\bar1}\mapsto\ket{e_{-i}}$, where $\ket{e_{\pm i}}$ are eigenstates of the AM operator $\xi$ with eigenvalues $\pm\lambda_i$. The MS interaction $U_{\text{MS}}(\Phi=\theta/4\lambda_i^2)=\exp({i\Phi\Xi^2})$ is then applied to yield the logical $Z$ rotation $\theta$, and $U^\dagger$ inverts the map. Inset: phase-space trajectories of the motional mode for each eigenstate of $\Xi=\xi\otimes I+I\otimes \xi$. The pair $\ket{e_{\pm i}e_{\pm i}}$ follows opposite loops (red and blue) that return to the origin after a gate duration of $t=\sqrt{\theta \pi}/\sqrt{2}\eta\lambda_i$.}
\end{figure}

The action of $U_{\text{MS}}$ on the logical states in Eq.~\ref{AECodeword} is set by the eigenvalues of the Hermitian AM operator $\xi$. The operator $\xi$ is constructed from the rank-1 and 2 STOs with $q=\pm1$ and therefore only couples nearest-neighbor Zeeman sublevels ($|\Delta m| = 1$). Such an operator has a set of eigenvalues which are symmetric about zero: its eigenstates $\ket{e_i}$, where $i \in \{-K, \dots, K\}$, satisfy $\xi\ket{e_i} = \lambda_i \ket{e_i}$ with $\lambda_{-i} = -\lambda_i$. 
Using this characteristic of the operator $\xi$, we can construct a logical two-qubit gate straightforwardly. For instance, to implement the two-qubit gate
\begin{align}
    R^L_{ZZ}(\theta)=e^{i\frac{\theta}{2} Z\otimes Z},
\end{align}
with $Z=\ket{\bar{1}}\bra{\bar{1}}-\ket{\bar{0}}\bra{\bar{0}}$, we first use universal control to map the logical states $\ket{\bar0}$ and $\ket{\bar1}$ onto $\ket{e_{\pm i}}$ for any $i\neq0$, then apply the generalized MS interaction, and finally undo the mapping. A key advantage of this approach is that a logical two-qubit gate reduces to the same simple structure as its physical counterpart. Beyond this logical two-qubit gate, by alternating $N$ single-qudit operations with MS interactions, one can synthesize arbitrary two-qudit operations~\cite{Brennen2005,Omanakuttan2023} (see Fig.~\ref{TwoQuditCircuitDiagram}). Furthermore, this generalized MS entangling interaction combined with universal single-qudit control may enable fault-tolerant gates~\cite{Omanakuttan2024,Onggadinata2026}, which we leave to future work.

\section{Discussion}
In summary, we introduce and numerically investigate a scalable, universal control protocol for implementing arbitrary $SU(d)$ operations within an AM manifold. The protocol employs phase-modulated two-photon stimulated Raman transitions that provide the rank-1 and rank-2 spherical tensor operators required for control universality. We show high-fidelity arbitrary state preparation and unitary transformations in representative integer- and half-integer-spin manifolds, with errors limited by photon scattering due to the Raman fields. The Raman configurations also offer complementary capabilities beyond those demonstrated here. The two independent phases of the $SPS$ configuration may enable simultaneous control of multiple hyperfine manifolds, such as those in the $D$ levels of $^{176}\text{Lu}^+$, $^{137}\text{Ba}^+$, and $^{43}\text{Ca}^+$~\cite{Qichen2026,Shi2026,Benhelm2007}, while the $|\Delta m|=2$ selectivity of the $SS$ configuration may suit encodings of multiple qubits in subspaces that remain unmixed under $SS$~\cite{Omanakuttan2021,Aydin2025,Low2026}.

We further show that the Raman transition can generate native two-qudit entangling operations. The resulting generalized MS interaction acts on the full AM manifold and, combined with universal single-qudit control, yields a logical two-qubit gate sequence that applies to any encoding within the manifold. More broadly, the generalized MS interaction together with single-qudit control provides universal two-qudit operations. 
Beyond the preparation and manipulation of the logical qubit encodings demonstrated here, these capabilities may also benefit quantum logic readout of logical qubits via a co-trapped ancilla~\cite{Tan2015}, hybrid qudit–oscillator computation in trapped ions~\cite{Heris2026}, and the encoding and recovery operations required for correction of logical qubits encoded in large AM manifolds~\cite{Furey2024}.

\begin{acknowledgments}
We thank Jai Karandikar, Dhruv Hariharan, Gabriel Gregory, Katya Fouka and Bas Gerritsen for helpful discussions. We were supported by the U.S.\ Office of Naval Research Global (N62909-24-1-2083), the U.S.\ Air Force Office of Scientific Research (FA2386-23-1-4062), the Australian Research Council (FT220100359, DP260104144), the University of Sydney Postgraduate Award scholarship (HKC, VGM), the Sydney Quantum Academy (PN), the Australian Government Research Training Program (FS), the Sydney Horizon Fellowship (TRT) and H.\ and A.\ Harley.
\end{acknowledgments}
\newpage
\appendix

\section{Rabi Frequency Decomposition into Spherical Tensors}\label{RabiFrequencyDecompositionIntoSphericalTensors}

We decompose the two-photon Rabi frequency $ \Omega_m^{\nu,\mu}$ in Eq.~\ref{Two-PhotonRabiFrequencyEquation} into STOs up to rank-$k=2$ following Refs.~\cite{Deutsch2010,Steck,LeKien2013}. We begin by decomposing the two-photon Rabi frequency into CG coefficients, Wigner-6$j$ symbols, and reduced matrix elements. We find
\begin{widetext}
\begin{align}
    \Omega_m^{\nu,\mu}&= (-1)^\nu\frac{E_\mu E_\nu}{4\bar{\Delta}_{\mu,\nu}}\sum_{m'}\langle K,m{+}q|d_{-\nu}|K',m'\rangle\langle K',m'|d_{\mu}|K,m \rangle \\
    &=(-1)^\nu\frac{E_\mu E_\nu}{4\bar{\Delta}_{\mu,\nu}}\langle K,m{+}q|\sum_{k=0}^{2}\sum_{q=-k}^{k}C^{k,q}_{1,-\nu,1,\mu}\Big(\sum_{m'}\mathbf{d}|K',m'\rangle\langle K',m'|\mathbf{d}\Big)_q^{(k)}|K,m\rangle  \\ 
    &=(-1)^\nu\frac{E_\mu E_\nu}{4\bar{\Delta}_{\mu,\nu}}\sum_{k=0}^{2}\sum_{q=-k}^{k}C^{k,q}_{1,-\nu,1,\mu} C^{K,m{+}q}_{K,m,k,q}\langle K,m{+}q||\Big(\sum_{m'}\mathbf{d}|K',m'\rangle\langle K',m'|\mathbf{d}\Big)^{(k)}||K,m\rangle \\
    &=(-1)^\nu\frac{E_\mu E_\nu}{4\bar{\Delta}_{\mu,\nu}}\sum_{k=0}^{2}\sum_{q=-k}^{k}C^{k,q}_{1,-\nu,1,\mu} C^{K,m{+}q}_{K,m,k,q}(-1)^{k+2K}\sqrt{(2K'+1)(2k+1)}\left\{
\begin{matrix}
1 & 1 & k \\
K & K & K'
\end{matrix} \right\}\langle K||\mathbf{d}||K'\rangle\langle K'||\mathbf{d}||K\rangle \\
&=(-1)^\nu\frac{E_\mu E_\nu}{4\bar{\Delta}_{\mu,\nu}}\sum_{k=0}^{2}\sum_{q=-k}^{k}C^{k,q}_{1,-\nu,1,\mu} C^{K,m{+}q}_{K,m,k,q}(-1)^{k+K+K'}\sqrt{(2K+1)(2k+1)}\left\{
\begin{matrix}
1 & 1 & k \\
K & K & K'
\end{matrix} \right\}|\langle K||\mathbf{d}||K'\rangle|^2. 
\end{align}
In the second line, we use the property associated with the direct product of STOs in the first equality of Eq.~\ref{DirectProductSTO}; in the third line, we use the Wigner-Eckart theorem; in the fourth line, we use the property of decomposing reduced matrix elements of operator products into products of reduced matrix elements, i.e.,
\begin{align}
    \langle J||\mathbf{A}^{(k)}||J'\rangle &=
    \langle J||\mathbf{B}^{(k_1)}\mathbf{C}^{(k_2)}||J'\rangle \\
    &=(-1)^{k+J+J'}\sum_{J''}\sqrt{(2J''+1)(2k+1)}\left\{
\begin{matrix}
k_1 & k_2 & k \\
J' & J & J''
\end{matrix} \right\}\langle J||\mathbf{B}^{(k_1)}||J''\rangle\langle J''||\mathbf{C}^{(k_2)}||J'\rangle. 
\end{align}
In the last line, we use the property associated with inverting the reduced matrix element, i.e.
\begin{align}
    \langle K'||\mathbf{d}||K\rangle=(-1)^{K'-K}\sqrt{\frac{2K+1}{2K'+1}}\langle K||\mathbf{d}||K'\rangle^*.
\end{align}
For atoms without nuclear spin $(K=J)$, the decomposition becomes
\begin{align}
    \Omega^{\nu,\mu}_m=(-1)^\nu\frac{E_\mu E_\nu}{4\bar{\Delta}_{\mu,\nu}}\sum_{k=0}^{2}\sum_{q=-k}^{k}C^{k,q}_{1,-\nu,1,\mu} C^{J,m{+}q}_{J,m,k,q}(-1)^{k+J+J'}\sqrt{(2J+1)(2k+1)}\left\{
\begin{matrix}
1 & 1 & k \\
J & J & J'
\end{matrix} \right\}|\langle J||\mathbf{d}||J'\rangle|^2,
\end{align}
whereas for atoms nuclear spin $(K=F)$ the decomposition is
\begin{align}
   \Omega_m^{\nu,\mu}&=(-1)^\nu\frac{E_\mu E_\nu}{4\bar{\Delta}_{\mu,\nu}}\sum_{k=0}^{2}\sum_{q=-k}^{k}C^{k,q}_{1,-\nu,1,\mu} C^{F,m{+}q}_{F,m,k,q}(-1)^{k+F+F'}\sqrt{(2F+1)(2k+1)}\left\{
\begin{matrix}
1 & 1 & k \\
F & F & F'
\end{matrix} \right\}|\langle F||\mathbf{d}||F'\rangle|^2 \\
& = (-1)^\nu\frac{E_\mu E_\nu}{4\bar{\Delta}_{\mu,\nu}}\sum_{k=0}^{2}\sum_{q=-k}^{k}C^{k,q}_{1,-\nu,1,\mu} C^{F,m{+}q}_{F,m,k,q}(-1)^{k+F+F'}(2F'+1)(2J+1)\sqrt{(2F+1)(2k+1)} \notag \\ &\quad \times\left\{
\begin{matrix}
1 & 1 & k \\
F & F & F'
\end{matrix} \right\}\left\{
\begin{matrix}
J & J' & 1 \\
F' & F & I
\end{matrix} \right\}^2|\langle J||\mathbf{d}||J'\rangle|^2.
\end{align}
In addition, for an atom with nuclear spin, the detuning $\bar{\Delta}_{\mu,\nu}$ of the Raman light fields from an excited AM $J'$ manifold is typically much larger than the hyperfine splitting between the individual $F'$ manifolds. This way we can treat the detuning $\bar{\Delta}_{\mu,\nu}$ to be independent of $F'$ and take the sum across coupling to all $F'$ within the same $J'$ excited AM manifold. Thus, in the limit that the detuning is much larger than the hyperfine splittings, we obtain
\begin{align}
           \sum_{F'} \Omega_m^{\nu,\mu}
           &=(-1)^\nu\frac{E_\mu E_\nu}{4\bar{\Delta}_{\mu,\nu}}\sum_{F'}\sum_{k=0}^{2}\sum_{q=-k}^{k}C^{k,q}_{1,-\nu,1,\mu} C^{F,m{+}q}_{F,m,k,q}(-1)^{k+F+F'}(2F'+1)(2J+1)\sqrt{(2F+1)(2k+1)} \notag \\ &\quad \times \left\{
\begin{matrix}
1 & 1 & k \\
F & F & F'
\end{matrix} \right\}\left\{
\begin{matrix}
J & J' & 1 \\
F' & F & I
\end{matrix} \right\}^2|\langle J||\mathbf{d}||J'\rangle|^2 \\
&=(-1)^\nu\frac{E_\mu E_\nu}{4\bar{\Delta}_{\mu,\nu}}\sum_{k=0}^{2}\sum_{q=-k}^{k}C^{k,q}_{1,-\nu,1,\mu} C^{F,m{+}q}_{F,m,k,q}(-1)^{-2J-J'-F-I}(2J+1)\sqrt{(2F+1)(2k+1)} \notag \\ &\quad \times\left\{
\begin{matrix}
1 & 1 & k \\
J & J & J'
\end{matrix} \right\}\left\{
\begin{matrix}
J & J & k \\
F & F & I
\end{matrix} \right\}|\langle J||\mathbf{d}||J'\rangle|^2
    \end{align}
where we use the property 
\begin{align}
    \sum_{F'}(-1)^{F'}(2F'+1)\left\{
\begin{matrix}
1 & 1 & k \\
F & F & F'
\end{matrix} \right\}\left\{
\begin{matrix}
J & J' & 1 \\
F' & F & I
\end{matrix} \right\}^2=(-1)^{-(2J+J'+2F+I+k)}\left\{
\begin{matrix}
1 & 1 & k \\
J & J & J'
\end{matrix} \right\} \left\{
\begin{matrix}
J & J & k \\
F & F & I
\end{matrix} \right\}.
\end{align}
Finally, we can write the two-photon Rabi frequency between states $\ket{K,m}$ and $\ket{K,m{+}q}$ as 
\begin{align}
\sum_{m} \Omega_m^{\nu,\mu}\ket{K,m{+}q}\bra{K,m} =
   (-1)^\nu\frac{E_\mu E_\nu}{4\bar{\Delta}_{\mu,\nu}}\sum_{k=0}^2C^{k,q}_{1,-\nu,1,\mu}A^{(k)}(K) T^{(k)}_q(K),
\end{align}
using the property $C^{K,m{+}q}_{K,m,k,q}=(-1)^kC^{K,m{+}q}_{k,q,K,m}$ we have for atoms with hyperfine AM $(K=F)$ the coefficient
\begin{align}
    A^{(k)}(F)&=(-1)^{-F-2J-J'-I+k}(2F+1)(2J+1)\left\{
\begin{matrix}
1 & 1 & k \\
J & J & J'
\end{matrix} \right\}\left\{
\begin{matrix}
J & J & k \\
F & F & I
\end{matrix} \right\}|\langle J||\mathbf{d}||J'\rangle|^2,
\end{align}
and for atoms without nuclear spin $(K=J)$
\begin{align}
    A^{(k)}(J)&=(-1)^{2k+J+J'}(2J+1)\left\{
\begin{matrix}
1 & 1 & k \\
J & J & J'
\end{matrix} \right\}|\langle J||\mathbf{d}||J'\rangle|^2 .
\end{align}
\end{widetext}

\section{Control Universality of the $SS$ Raman Configuration in the AM Subspace}\label{Control Universality}
We begin by reviewing the principle of universal control of an AM manifold introduced in Ref.~\cite{Merkel2008}, and then show that the $SS$ configuration provides universal control over the AM subspace closed under its $|\Delta m|=2$ couplings.

Given a set of operators such as $\{K_x, K_y, \hat{O}\}$, the unitary evolution is any element of their Lie algebra, i.e., all linear combinations and nested commutators of these operators can be generated through Trotterization~\cite{Lloyd1995}. Universal control over an AM manifold of spin $K$ requires this algebra to contain all STOs up to rank $2K$. This holds if $\hat{O}$ has nonzero overlap with some rank-$2$ STO, $\mathrm{Tr}(\hat{O}^\dagger T_q^{(2)})\neq0$. To see this, we consider the commutator of two general STOs. The direct product between two STOs $ T^{(k_1)}_{q_1}$ and $ T^{(k_2)}_{q_2}$ is given by
\begin{align}
    T^{(k_1)}_{q_1}T^{(k_2)}_{q_2}
    &=\sum_{L=|k_1-k_2|}^{k_1+k_2}\sum_{q=-L}^L\,
     C^{L,q}_{k_1,q_1,k_2,q_2}
    (T^{(k_1)}T^{(k_2)})^{(L)}_q \notag \\ 
    &=
    \sum_{L=|k_1-k_2|}^{k_1+k_2}\sum_{q=-L}^L\,
     \beta_L C^{L,q}_{k_1,q_1,k_2,q_2}
    T^{(L)}_{q}.\label{DirectProductSTO}
\end{align}
where the CG coefficient $C^{L,q}_{k_1,q_1,k_2,q_2}$ is nonzero provided $q = q_1+q_2$ and $|k_1-k_2|\leq L\leq k_1+k_2$, and $\beta_L = (-1)^{2K+k_1+k_2}\sqrt{(2k_1+1)(2k_2+1)}\left\{
\begin{matrix}
k_1 & k_2 & L \\
K & K & K
\end{matrix} \right\}$ is nonzero only for $|k_1-k_2|\le L\le\min(k_1+k_2,2K)$. Therefore, we can write the commutator of two spherical tensor operators as
\begin{align}
    [T^{(k_1)}_{q_1}, T^{(k_2)}_{q_2}] &=  \sum_L \beta_LC^{L,q_1+q_2}_{k_1,q_1,k_2,q_2} T^{(L)}_{q_1 + q_2} \notag \\ &\quad -\sum_L \beta_L C^{L,q_1+q_2}_{k_2,q_2,k_1,q_1} T^{(L)}_{q_1 + q_2} \notag \\ 
    &= \sum_L \beta_LC^{L,q_1+q_2}_{k_1,q_1,k_2,q_2} T^{(L)}_{q_1 + q_2} \times \notag \\ &\quad\big(1-(-1)^{k_1+k_2-L}\big), \label{SphericalTensorCommutationRelation}
\end{align}
where we have used the symmetry relation of CG coefficients $ C^{L,q_1+q_2}_{k_2,q_2,k_1,q_1} =  (-1)^{k_1+k_2-L}C^{L,q_1+q_2}_{k_1,q_1,k_2,q_2}$ and $\beta_L$ is constant under $k_1\leftrightarrow k_2$ exchange. The commutator vanishes unless $k_1+k_2-L$ is odd, $|k_1-k_2|\leq L\leq k_1+k_2$ and $L\leq 2K$.

We now consider the commutators generated by $\{K_x, K_y, \hat{O}\}$. Writing the AM raising and lowering operators as $K_{\pm}=K_x\pm iK_y\propto T^{(1)}_{\pm1}$ and taking $\hat{O}\propto T^{(2)}_{q_1}$, we have
\begin{align}
    \left[K_{\pm}, T^{(k)}_{q}\right] &\propto T^{(k)}_{q\pm1}, \label{commutationrelationJplusJminusTk} \\
    \left[\hat{O}, T^{(k)}_{q_2}\right]&\propto \beta_{k+1}C^{k+1,q_1+q_2}_{2,q_1,k,q_2}\,T^{(k+1)}_{q_1+q_2} + \notag \\ &\quad \beta_{k-1}C^{k-1,q_1+q_2}_{2,q_1,k,q_2}\,T^{(k-1)}_{q_1+q_2}. \label{commutationrelationT2withTk}
\end{align}
By Eq.~\ref{commutationrelationJplusJminusTk}, repeated commutation of $T^{(k)}_q$ with $K_{\pm}$ generates all STOs of rank $k$. By Eq.~\ref{commutationrelationT2withTk}, commutation with $\hat{O}$ raises the rank by one, while the rank-$(k-1)$ component is already in the algebra. Starting from $K_{\pm}$ and iterating, $\{K_x, K_y, \hat{O}\}$ therefore generates all STOs up to rank $2K$, giving universal control over an AM manifold of spin $K$.

We now show that the $SS$ configuration is universal within each subspace closed under $|\Delta m|=2$ couplings. Because $SS$ control couples only states $\ket{m}$ with $|\Delta m|=2$, the manifold splits into two Zeeman subspaces and it does not mix, for instance, the even-$m$ and odd-$m$ levels of an integer-spin manifold. Universal control over these subspaces requires the Lie algebra to contain all STOs with even $q$. The $SS$ configuration provides the operators $\{T^{(2)}_{\pm2}, T^{(2)}_0\}$. By Eq.~\ref{commutationrelationT2withTk}, commuting two rank-$2$ STOs yields the odd-rank STOs $T^{(1)}_0$ and $T^{(3)}_{q}$ with even $q$. Commuting these with the rank-$2$ generators then yields even-rank STOs, again with even $q$. Iterating this alternation between odd and even ranks generates all even-$q$ STOs up to rank $2K$, giving universal control within each subspace.

\section{Phase Modulation and Quantum Optimal Control}\label{PhaseModulationOptimalControl}
We describe the methodology for obtaining the phase-modulated waveforms for state preparation, unitary evolution, and the single-qubit operations for the error-correcting codewords.

We discretize the effective Hamiltonian into $N$ piece-wise-constant (pwc) segments, where the $k$-th segment spans the time interval $[t_k, t_{k+1}]$ with duration $\delta t_k = t_{k+1}-t_k$, for $k=1,\dots,N$, with $t_1=0$ and $t_{N+1}=T$ the total pulse duration. The Hamiltonian in each segment is defined by a vector of phases, i.e.,
$H^{SPS}_{\text{eff}}(\boldsymbol{\phi}(t_k))$, where $\boldsymbol{\phi}(t)=\left(\phi_{+1,0}(t),\phi_{0,-1}(t)\right)$ is held constant over the segment. The phase of the $|\Delta m|=2$ term is not an independent control but is fixed by $\phi_{+1,-1}(t)=\phi_{+1,0}(t)+\phi_{0,-1}(t)$. For the $SP$ and $SS$ configurations, obtained from $H^{SPS}_{\text{eff}}$ by setting $E_{+1}=0$ and $E_0=0$ (Sec.~\ref{examples section}), $\boldsymbol{\phi}(t)$ reduces to the single phase $\phi_{0,-1}(t)$ and $\phi_{+1,-1}(t)$, respectively. This constructs a unitary operator given by
\begin{align}
    U_c=\prod_{k=N}^{1} e^{-i H^{SPS}_\text{eff}(\boldsymbol{\phi}(t_k)) \delta t_k}
    \label{UnitaryEvolutionEquation}
\end{align}
to perform any operation in $SU(d)$ by numerically optimizing the phases $\boldsymbol{\phi}(t)$ using a gradient ascent algorithm from the Q-CTRL Boulder Opal package~\cite{Ball2021,Bentley2020}.

For state preparation, we find the optimal phases by minimizing the infidelity 
\begin{align}
    \mathcal{I}_{\text{s}} &= 1-|\bra{\psi_\text{tar}}U_c\ket{\psi_{\text{ini}}}|^2,
\end{align}
where $\psi_\text{tar}$ and $\psi_\text{ini}$ denotes the target and initial state. 

For unitary evolution, we find the optimal phases by minimizing the cost 
\begin{align}
     C&=\mathcal{I}_{\text{u}} + \varepsilon\frac{T}{T_{\text{max}}}\\ &= 1-|\frac{\text{Tr}\big(U_{\text{tar}}^\dagger U_c\big)}{\text{Tr}\big(U_{\text{tar}}^\dagger U_{\text{tar}}\big)} |^2+\varepsilon\frac{T}{T_{\text{max}}},
\end{align}
where $U_{\text{tar}}$ is the target unitary operation, $T=N\delta t_k$ is the total waveform duration, $T_{\text{max}}$ is the maximum allowed duration and $\varepsilon$ is a penalty on duration optimization. 
We heuristically set $N=4*(2d-2)$ optimizable pwc segments for state preparation and $N=2*(d^2-1)$ for unitary evolutions, where $d$ is the dimension of the accessible AM Hilbert space. For state preparation, the duration $T$ is fixed in the optimization;
whereas for unitary evolution, $T$ is also optimizable. 

We set the penalty $\varepsilon=10^{-6}$ such that the optimizer minimizes the duration $T$ while ensuring $\mathcal{I}_\text{u}\leq \varepsilon$, and the maximum duration $T_{\text{max}}$ is chosen such that the optimization can converge in a reasonable amount of wall-clock time.

To implement a logical single-qubit gate, we optimize the control phases such that the resulting unitary $U_c$ maps each of the six logical Pauli eigenstates to the state produced under the action of the target gate. In terms of the logical codewords $\ket{\bar{0}}$ and $\ket{\bar{1}}$, these eigenstates are
\begin{align}
    \ket{+Z}_L &= \ket{\bar{1}}, \notag \\
    \ket{-Z}_L &= \ket{\bar{0}}, \notag \\
    \ket{\pm X}_L &= \tfrac{1}{\sqrt{2}}\left(\ket{\bar{0}} \pm \ket{\bar{1}}\right), \notag \\
    \ket{\pm Y}_L &= \tfrac{1}{\sqrt{2}}\left(\ket{\bar{0}} \mp i\ket{\bar{1}}\right),
\end{align}
which are the eigenstates of the logical Pauli operators $Z=\ket{\bar1}\bra{\bar1}-\ket{\bar0}\bra{\bar0}$, $X=\ket{\bar0}\bra{\bar1}+\ket{\bar1}\bra{\bar0}$, and $Y=iXZ$. The target gates are the logical Hadamard $H_\text{AE}=(X+Z)/\sqrt{2}$ and the logical $T$ gate $T_\text{AE}=\ket{\bar1}\bra{\bar1}+e^{i\pi/4}\ket{\bar0}\bra{\bar0}$. For example, $H_\text{AE}$ maps $\ket{+X}_L$ to $\ket{+Z}_L$, so the optimized $U_c$ is required to do the same. Because these state-to-state mappings constrain $U_c$ only on the code space, leaving its action on the rest of the manifold free, they simplify the search for optimal control solutions~\cite{Merkel2009,Matsos2024,Matsos2025}. For each target gate $U_\text{tar}\in\{H_\text{AE},T_\text{AE}\}$, we find the optimal phases by minimizing the state-averaged infidelity
\begin{align}
    \mathcal{I}=1-\frac{1}{6}\sum_{i} \left|\bra{i}U_\text{tar}^\dagger U_c\ket{i}\right|^2, \quad i\in\{\pm X,\pm Y,\pm Z\},
    \label{HandTStatemap}
\end{align}
following Refs.~\cite{Matsos2024,Matsos2025,Millican2025}. Using the resulting $U_c$, we then evaluate the average gate fidelity of each logical single-qubit gate following Ref.~\cite{Matsos2025}.

\section{State Preparation Results and Calculation of Photon Scattering Errors in the Raman Transition}\label{PhotonScatterEstimate}

State preparation infidelities and the associated photon scattering errors for the spin-cat states, BO and AE codewords, are reported in Table~\ref{UniversalControlFidelities_stateprep}. In the absence of photon scatter errors, state preparation infidelities of our protocol for both the $^{40}\text{Ca}^+$ and $^{176}\text{Lu}^+$ ions are below $10^{-6}$. 

We give a pessimistic estimate of the intrinsic infidelity of Raman transitions by considering the following photon scattering rate~\cite{Ozeri2005,Ozeri2007,Wineland2003}
\begin{align}
    R&\approx\sum_m\sum_{K'}\sum_q\sum_{m'}\frac{P_m\gamma_{K'|\langle K',m'|d_qE_q   |K,m\rangle|^2}}{4\hbar^2\Delta_{K'}^2}.
\end{align}
The scattering rate is obtained by multiplying the decay rate $\gamma_{K'}$ of each excited state $\ket{K',m'}$ by its off-resonantly excited population, $P_m|\langle K',m'|d_qE_q|K,m\rangle|^2/\Delta_{K'}^2$, where $P_m$ is the population occupying $\ket{K,m}$ and $q$ denotes the field polarization. Since the optimized pulses distribute population across the accessible sublevels, we approximate $P_m \approx 1/d$, where $d$ is the dimension of the subspace accessible to each Raman configuration given in Sec.~\ref{Arbitrary State Preparation and Unitary Evolution}.

We compute the photon-scattering error by multiplying the scattering rate by the duration of the operation. The estimated infidelity is in the order of $10^{-5}$ and $10^{-3}$ for the $^{40}\text{Ca}^+$ and $^{176}\text{Lu}^+$ ions, respectively. A more precise infidelity estimate can be obtained by considering the master equation treatment along with different photon scatter pathways~\cite{Deutsch2010,Moore2023}.

\begin{table*}
\centering
\setlength{\tabcolsep}{8pt}
\renewcommand{\arraystretch}{0.95}
\begin{tabular}{lccccc}
\specialrule{1.2pt}{0pt}{1pt}
\specialrule{0.4pt}{1pt}{3pt}
Target & $\log_{10}$(Infidelity) & Duration ($\Omega_\mathbf{K} T/2\pi$) &  $\log_{10}$(Photon scatter error) & Raman config. & Ion\\
\specialrule{0.4pt}{2pt}{2pt}
$\ket{\bar{0}}_{\text{BO}}$ & $-8.53$ & $0.43$ &$-5.45$ & SPS &$^{40}\text{Ca}^+$ \\
$\ket{\bar{1}}_{\text{BO}}$ & $-8.26$ & $0.43$ &$-5.45$ & SPS &$^{40}\text{Ca}^+$\\
$\ket{\text{cat}}_{5/2}$ & $-8.45$ & $0.43$ &$-5.45$ & SPS &$^{40}\text{Ca}^+$ \\
$\ket{\bar{0}}_{AE}$ & $-8.22$ & $3.43$ &$-3.52$ & SPS &$^{176}\text{Lu}^+$ \\
$\ket{\bar{0}}_{AE}$ & $-6.54$ & $6.84$ &$-3.22$ & SP &$^{176}\text{Lu}^+$\\
$\ket{\bar{0}}_{AE}$ & $-8.41$ & $3.43$ &$-3.51$ & SS &$^{176}\text{Lu}^+$ \\
$\ket{\bar{1}}_{AE}$ & $-8.13$ & $3.43$ &$-3.52$& SPS &$^{176}\text{Lu}^+$ \\
$\ket{\bar{1}}_{AE}$ & $-6.34$ & $6.84$ &$-3.22$ & SP &$^{176}\text{Lu}^+$ \\
$\ket{\bar{1}}_{AE}$ & $-8.44$ & $3.43$ &$-3.51$ & SS &$^{176}\text{Lu}^+$ \\
$\ket{\text{cat}}_{8}$ & $-8.11$ & $3.43$ &$-3.52$ & SPS &$^{176}\text{Lu}^+$ \\
$\ket{\text{cat}}_{8}$ & $-7.32$ & $6.84$ &$-3.22$ & SP &$^{176}\text{Lu}^+$ \\
$\ket{\text{cat}}_{8}$ & $-8.30$ & $3.43$ &$-3.51$ & SS &$^{176}\text{Lu}^+$ \\
\specialrule{0.4pt}{2pt}{1pt}
\specialrule{1.2pt}{1pt}{0pt}
\end{tabular}
\caption{\label{UniversalControlFidelities_stateprep}Noiseless infidelities, pulse durations, and estimated photon scattering errors for state preparation in $^{176}\text{Lu}^+$ and $^{40}\text{Ca}^+$ using different Raman configurations.}
\end{table*}

\bibliography{references_updated}

\begin{thebibliography}{72}%
\makeatletter
\providecommand \@ifxundefined [1]{%
 \@ifx{#1\undefined}
}%
\providecommand \@ifnum [1]{%
 \ifnum #1\expandafter \@firstoftwo
 \else \expandafter \@secondoftwo
 \fi
}%
\providecommand \@ifx [1]{%
 \ifx #1\expandafter \@firstoftwo
 \else \expandafter \@secondoftwo
 \fi
}%
\providecommand \natexlab [1]{#1}%
\providecommand \enquote  [1]{``#1''}%
\providecommand \bibnamefont  [1]{#1}%
\providecommand \bibfnamefont [1]{#1}%
\providecommand \citenamefont [1]{#1}%
\providecommand \href@noop [0]{\@secondoftwo}%
\providecommand \href [0]{\begingroup \@sanitize@url \@href}%
\providecommand \@href[1]{\@@startlink{#1}\@@href}%
\providecommand \@@href[1]{\endgroup#1\@@endlink}%
\providecommand \@sanitize@url [0]{\catcode `\\12\catcode `\$12\catcode `\&12\catcode `\#12\catcode `\^12\catcode `\_12\catcode `\%12\relax}%
\providecommand \@@startlink[1]{}%
\providecommand \@@endlink[0]{}%
\providecommand \url  [0]{\begingroup\@sanitize@url \@url }%
\providecommand \@url [1]{\endgroup\@href {#1}{\urlprefix }}%
\providecommand \urlprefix  [0]{URL }%
\providecommand \Eprint [0]{\href }%
\providecommand \doibase [0]{https://doi.org/}%
\providecommand \selectlanguage [0]{\@gobble}%
\providecommand \bibinfo  [0]{\@secondoftwo}%
\providecommand \bibfield  [0]{\@secondoftwo}%
\providecommand \translation [1]{[#1]}%
\providecommand \BibitemOpen [0]{}%
\providecommand \bibitemStop [0]{}%
\providecommand \bibitemNoStop [0]{.\EOS\space}%
\providecommand \EOS [0]{\spacefactor3000\relax}%
\providecommand \BibitemShut  [1]{\csname bibitem#1\endcsname}%
\let\auto@bib@innerbib\@empty
\bibitem [{\citenamefont {Meth}\ \emph {et~al.}(2025)\citenamefont {Meth}, \citenamefont {Zhang}, \citenamefont {Haase}, \citenamefont {Edmunds}, \citenamefont {Postler}, \citenamefont {Jena}, \citenamefont {Steiner}, \citenamefont {Dellantonio}, \citenamefont {Blatt}, \citenamefont {Zoller}, \citenamefont {Monz}, \citenamefont {Schindler}, \citenamefont {Muschik},\ and\ \citenamefont {Ringbauer}}]{Meth2025}%
  \BibitemOpen
  \bibfield  {author} {\bibinfo {author} {\bibfnamefont {M.}~\bibnamefont {Meth}}, \bibinfo {author} {\bibfnamefont {J.}~\bibnamefont {Zhang}}, \bibinfo {author} {\bibfnamefont {J.~F.}\ \bibnamefont {Haase}}, \bibinfo {author} {\bibfnamefont {C.}~\bibnamefont {Edmunds}}, \bibinfo {author} {\bibfnamefont {L.}~\bibnamefont {Postler}}, \bibinfo {author} {\bibfnamefont {A.~J.}\ \bibnamefont {Jena}}, \bibinfo {author} {\bibfnamefont {A.}~\bibnamefont {Steiner}}, \bibinfo {author} {\bibfnamefont {L.}~\bibnamefont {Dellantonio}}, \bibinfo {author} {\bibfnamefont {R.}~\bibnamefont {Blatt}}, \bibinfo {author} {\bibfnamefont {P.}~\bibnamefont {Zoller}}, \bibinfo {author} {\bibfnamefont {T.}~\bibnamefont {Monz}}, \bibinfo {author} {\bibfnamefont {P.}~\bibnamefont {Schindler}}, \bibinfo {author} {\bibfnamefont {C.}~\bibnamefont {Muschik}},\ and\ \bibinfo {author} {\bibfnamefont {M.}~\bibnamefont {Ringbauer}},\ }\bibfield  {title} {\bibinfo {title} {Simulating two-dimensional lattice gauge theories on a qudit quantum
  computer},\ }\href@noop {} {\bibfield  {journal} {\bibinfo  {journal} {Nature Physics}\ }\textbf {\bibinfo {volume} {21}},\ \bibinfo {pages} {570} (\bibinfo {year} {2025})}\BibitemShut {NoStop}%
\bibitem [{\citenamefont {John}\ \emph {et~al.}(2026)\citenamefont {John}, \citenamefont {Pareek}, \citenamefont {Tirler}, \citenamefont {Gollerthan}, \citenamefont {Meth}, \citenamefont {Gerster}, \citenamefont {Zoller}, \citenamefont {Gonz{\'a}lez-Cuadra}, \citenamefont {Zache},\ and\ \citenamefont {Ringbauer}}]{John2026}%
  \BibitemOpen
  \bibfield  {author} {\bibinfo {author} {\bibfnamefont {M.}~\bibnamefont {John}}, \bibinfo {author} {\bibfnamefont {K.}~\bibnamefont {Pareek}}, \bibinfo {author} {\bibfnamefont {P.}~\bibnamefont {Tirler}}, \bibinfo {author} {\bibfnamefont {T.}~\bibnamefont {Gollerthan}}, \bibinfo {author} {\bibfnamefont {M.}~\bibnamefont {Meth}}, \bibinfo {author} {\bibfnamefont {L.}~\bibnamefont {Gerster}}, \bibinfo {author} {\bibfnamefont {P.}~\bibnamefont {Zoller}}, \bibinfo {author} {\bibfnamefont {D.}~\bibnamefont {Gonz{\'a}lez-Cuadra}}, \bibinfo {author} {\bibfnamefont {T.~V.}\ \bibnamefont {Zache}},\ and\ \bibinfo {author} {\bibfnamefont {M.}~\bibnamefont {Ringbauer}},\ }\href@noop {} {\bibinfo {title} {Non-abelian string-breaking dynamics on a qudit quantum computer}} (\bibinfo {year} {2026}),\ \Eprint {https://arxiv.org/abs/2605.05841} {arXiv:2605.05841 [quant-ph]} \BibitemShut {NoStop}%
\bibitem [{\citenamefont {Champion}\ \emph {et~al.}(2025{\natexlab{a}})\citenamefont {Champion}, \citenamefont {Schwartz}, \citenamefont {Ijaz}, \citenamefont {Xu}, \citenamefont {Campbell}, \citenamefont {Landi},\ and\ \citenamefont {Blok}}]{champion2025lmg}%
  \BibitemOpen
  \bibfield  {author} {\bibinfo {author} {\bibfnamefont {E.}~\bibnamefont {Champion}}, \bibinfo {author} {\bibfnamefont {A.}~\bibnamefont {Schwartz}}, \bibinfo {author} {\bibfnamefont {M.~A.}\ \bibnamefont {Ijaz}}, \bibinfo {author} {\bibfnamefont {X.}~\bibnamefont {Xu}}, \bibinfo {author} {\bibfnamefont {S.}~\bibnamefont {Campbell}}, \bibinfo {author} {\bibfnamefont {G.~T.}\ \bibnamefont {Landi}},\ and\ \bibinfo {author} {\bibfnamefont {M.~S.}\ \bibnamefont {Blok}},\ }\href@noop {} {\bibinfo {title} {Analog quantum simulation of the {Lipkin-Meshkov-Glick} model in a transmon qudit}} (\bibinfo {year} {2025}{\natexlab{a}}),\ \Eprint {https://arxiv.org/abs/2512.05237} {arXiv:2512.05237 [quant-ph]} \BibitemShut {NoStop}%
\bibitem [{\citenamefont {Gottesman}\ \emph {et~al.}(2001)\citenamefont {Gottesman}, \citenamefont {Kitaev},\ and\ \citenamefont {Preskill}}]{Gottesman2001}%
  \BibitemOpen
  \bibfield  {author} {\bibinfo {author} {\bibfnamefont {D.}~\bibnamefont {Gottesman}}, \bibinfo {author} {\bibfnamefont {A.}~\bibnamefont {Kitaev}},\ and\ \bibinfo {author} {\bibfnamefont {J.}~\bibnamefont {Preskill}},\ }\bibfield  {title} {\bibinfo {title} {Encoding a qubit in an oscillator},\ }\href@noop {} {\bibfield  {journal} {\bibinfo  {journal} {Physical Review A}\ }\textbf {\bibinfo {volume} {64}},\ \bibinfo {pages} {012310} (\bibinfo {year} {2001})}\BibitemShut {NoStop}%
\bibitem [{\citenamefont {Wang}\ \emph {et~al.}(2020)\citenamefont {Wang}, \citenamefont {Hu}, \citenamefont {Sanders},\ and\ \citenamefont {Kais}}]{Wang2020}%
  \BibitemOpen
  \bibfield  {author} {\bibinfo {author} {\bibfnamefont {Y.}~\bibnamefont {Wang}}, \bibinfo {author} {\bibfnamefont {Z.}~\bibnamefont {Hu}}, \bibinfo {author} {\bibfnamefont {B.~C.}\ \bibnamefont {Sanders}},\ and\ \bibinfo {author} {\bibfnamefont {S.}~\bibnamefont {Kais}},\ }\bibfield  {title} {\bibinfo {title} {Qudits and high-dimensional quantum computing},\ }\href@noop {} {\bibfield  {journal} {\bibinfo  {journal} {Frontiers in Physics}\ }\textbf {\bibinfo {volume} {8}},\ \bibinfo {pages} {589504} (\bibinfo {year} {2020})}\BibitemShut {NoStop}%
\bibitem [{\citenamefont {Pirandola}\ \emph {et~al.}(2008)\citenamefont {Pirandola}, \citenamefont {Mancini}, \citenamefont {Braunstein},\ and\ \citenamefont {Vitali}}]{Pirandola2008}%
  \BibitemOpen
  \bibfield  {author} {\bibinfo {author} {\bibfnamefont {S.}~\bibnamefont {Pirandola}}, \bibinfo {author} {\bibfnamefont {S.}~\bibnamefont {Mancini}}, \bibinfo {author} {\bibfnamefont {S.~L.}\ \bibnamefont {Braunstein}},\ and\ \bibinfo {author} {\bibfnamefont {D.}~\bibnamefont {Vitali}},\ }\bibfield  {title} {\bibinfo {title} {Minimal qudit code for a qubit in the phase-damping channel},\ }\href@noop {} {\bibfield  {journal} {\bibinfo  {journal} {Physical Review A}\ }\textbf {\bibinfo {volume} {77}},\ \bibinfo {pages} {032309} (\bibinfo {year} {2008})}\BibitemShut {NoStop}%
\bibitem [{\citenamefont {Chiesa}\ \emph {et~al.}(2020)\citenamefont {Chiesa}, \citenamefont {Macaluso}, \citenamefont {Petiziol}, \citenamefont {Wimberger}, \citenamefont {Santini},\ and\ \citenamefont {Carretta}}]{Chiesa2020}%
  \BibitemOpen
  \bibfield  {author} {\bibinfo {author} {\bibfnamefont {A.}~\bibnamefont {Chiesa}}, \bibinfo {author} {\bibfnamefont {E.}~\bibnamefont {Macaluso}}, \bibinfo {author} {\bibfnamefont {F.}~\bibnamefont {Petiziol}}, \bibinfo {author} {\bibfnamefont {S.}~\bibnamefont {Wimberger}}, \bibinfo {author} {\bibfnamefont {P.}~\bibnamefont {Santini}},\ and\ \bibinfo {author} {\bibfnamefont {S.}~\bibnamefont {Carretta}},\ }\bibfield  {title} {\bibinfo {title} {Molecular nanomagnets as qubits with embedded quantum-error correction},\ }\href@noop {} {\bibfield  {journal} {\bibinfo  {journal} {The Journal of Physical Chemistry Letters}\ }\textbf {\bibinfo {volume} {11}},\ \bibinfo {pages} {8610} (\bibinfo {year} {2020})}\BibitemShut {NoStop}%
\bibitem [{\citenamefont {Albert}\ \emph {et~al.}(2020)\citenamefont {Albert}, \citenamefont {Covey},\ and\ \citenamefont {Preskill}}]{Albert2020}%
  \BibitemOpen
  \bibfield  {author} {\bibinfo {author} {\bibfnamefont {V.~V.}\ \bibnamefont {Albert}}, \bibinfo {author} {\bibfnamefont {J.~P.}\ \bibnamefont {Covey}},\ and\ \bibinfo {author} {\bibfnamefont {J.}~\bibnamefont {Preskill}},\ }\bibfield  {title} {\bibinfo {title} {Robust encoding of a qubit in a molecule},\ }\href@noop {} {\bibfield  {journal} {\bibinfo  {journal} {Physical Review X}\ }\textbf {\bibinfo {volume} {10}},\ \bibinfo {pages} {031050} (\bibinfo {year} {2020})}\BibitemShut {NoStop}%
\bibitem [{\citenamefont {Gross}(2021)}]{Gross2021}%
  \BibitemOpen
  \bibfield  {author} {\bibinfo {author} {\bibfnamefont {J.~A.}\ \bibnamefont {Gross}},\ }\bibfield  {title} {\bibinfo {title} {Designing codes around interactions: The case of a spin},\ }\href@noop {} {\bibfield  {journal} {\bibinfo  {journal} {Physical Review Letters}\ }\textbf {\bibinfo {volume} {127}},\ \bibinfo {pages} {010504} (\bibinfo {year} {2021})}\BibitemShut {NoStop}%
\bibitem [{\citenamefont {Gross}\ \emph {et~al.}(2024)\citenamefont {Gross}, \citenamefont {Godfrin}, \citenamefont {Blais},\ and\ \citenamefont {Dupont-Ferrier}}]{Gross2024}%
  \BibitemOpen
  \bibfield  {author} {\bibinfo {author} {\bibfnamefont {J.~A.}\ \bibnamefont {Gross}}, \bibinfo {author} {\bibfnamefont {C.}~\bibnamefont {Godfrin}}, \bibinfo {author} {\bibfnamefont {A.}~\bibnamefont {Blais}},\ and\ \bibinfo {author} {\bibfnamefont {E.}~\bibnamefont {Dupont-Ferrier}},\ }\bibfield  {title} {\bibinfo {title} {Hardware-efficient error-correcting codes for large nuclear spins},\ }\href@noop {} {\bibfield  {journal} {\bibinfo  {journal} {Physical Review Applied}\ }\textbf {\bibinfo {volume} {22}},\ \bibinfo {pages} {014006} (\bibinfo {year} {2024})}\BibitemShut {NoStop}%
\bibitem [{\citenamefont {Lim}\ \emph {et~al.}(2023)\citenamefont {Lim}, \citenamefont {Liu},\ and\ \citenamefont {Ardavan}}]{Lim2023}%
  \BibitemOpen
  \bibfield  {author} {\bibinfo {author} {\bibfnamefont {S.}~\bibnamefont {Lim}}, \bibinfo {author} {\bibfnamefont {J.}~\bibnamefont {Liu}},\ and\ \bibinfo {author} {\bibfnamefont {A.}~\bibnamefont {Ardavan}},\ }\bibfield  {title} {\bibinfo {title} {Fault-tolerant qubit encoding using a spin-7/2 qudit},\ }\href@noop {} {\bibfield  {journal} {\bibinfo  {journal} {Physical Review A}\ }\textbf {\bibinfo {volume} {108}},\ \bibinfo {pages} {062403} (\bibinfo {year} {2023})}\BibitemShut {NoStop}%
\bibitem [{\citenamefont {Lim}(2026)}]{Lim2026}%
  \BibitemOpen
  \bibfield  {author} {\bibinfo {author} {\bibfnamefont {S.}~\bibnamefont {Lim}},\ }\bibfield  {title} {\bibinfo {title} {Fault-tolerant encoding of logical qudits in spin systems},\ }\href@noop {} {\bibfield  {journal} {\bibinfo  {journal} {Physical Review A}\ }\textbf {\bibinfo {volume} {113}},\ \bibinfo {pages} {042439} (\bibinfo {year} {2026})}\BibitemShut {NoStop}%
\bibitem [{\citenamefont {Omanakuttan}\ \emph {et~al.}(2024)\citenamefont {Omanakuttan}, \citenamefont {Buchemmavari}, \citenamefont {Gross}, \citenamefont {Deutsch},\ and\ \citenamefont {Marvian}}]{Omanakuttan2024}%
  \BibitemOpen
  \bibfield  {author} {\bibinfo {author} {\bibfnamefont {S.}~\bibnamefont {Omanakuttan}}, \bibinfo {author} {\bibfnamefont {V.}~\bibnamefont {Buchemmavari}}, \bibinfo {author} {\bibfnamefont {J.~A.}\ \bibnamefont {Gross}}, \bibinfo {author} {\bibfnamefont {I.~H.}\ \bibnamefont {Deutsch}},\ and\ \bibinfo {author} {\bibfnamefont {M.}~\bibnamefont {Marvian}},\ }\bibfield  {title} {\bibinfo {title} {Fault-tolerant quantum computation using large spin-cat codes},\ }\href@noop {} {\bibfield  {journal} {\bibinfo  {journal} {PRX Quantum}\ }\textbf {\bibinfo {volume} {5}},\ \bibinfo {pages} {020355} (\bibinfo {year} {2024})}\BibitemShut {NoStop}%
\bibitem [{\citenamefont {Jain}\ \emph {et~al.}(2024)\citenamefont {Jain}, \citenamefont {Hudson}, \citenamefont {Campbell},\ and\ \citenamefont {Albert}}]{Jain2024}%
  \BibitemOpen
  \bibfield  {author} {\bibinfo {author} {\bibfnamefont {S.~P.}\ \bibnamefont {Jain}}, \bibinfo {author} {\bibfnamefont {E.~R.}\ \bibnamefont {Hudson}}, \bibinfo {author} {\bibfnamefont {W.~C.}\ \bibnamefont {Campbell}},\ and\ \bibinfo {author} {\bibfnamefont {V.~V.}\ \bibnamefont {Albert}},\ }\bibfield  {title} {\bibinfo {title} {Absorption-emission codes for atomic and molecular quantum information platforms},\ }\href@noop {} {\bibfield  {journal} {\bibinfo  {journal} {Physical Review Letters}\ }\textbf {\bibinfo {volume} {133}},\ \bibinfo {pages} {260601} (\bibinfo {year} {2024})}\BibitemShut {NoStop}%
\bibitem [{\citenamefont {Aydin}\ and\ \citenamefont {Barg}(2025)}]{Aydin2025}%
  \BibitemOpen
  \bibfield  {author} {\bibinfo {author} {\bibfnamefont {A.}~\bibnamefont {Aydin}}\ and\ \bibinfo {author} {\bibfnamefont {A.}~\bibnamefont {Barg}},\ }\bibfield  {title} {\bibinfo {title} {Class of codes correcting absorptions and emissions},\ }\href@noop {} {\bibfield  {journal} {\bibinfo  {journal} {Physical Review A}\ }\textbf {\bibinfo {volume} {111}},\ \bibinfo {pages} {022415} (\bibinfo {year} {2025})}\BibitemShut {NoStop}%
\bibitem [{\citenamefont {Furey}\ \emph {et~al.}(2024)\citenamefont {Furey}, \citenamefont {Wu}, \citenamefont {Isaza-Monsalve}, \citenamefont {Walser}, \citenamefont {Mattivi}, \citenamefont {Nardi},\ and\ \citenamefont {Schindler}}]{Furey2024}%
  \BibitemOpen
  \bibfield  {author} {\bibinfo {author} {\bibfnamefont {B.~J.}\ \bibnamefont {Furey}}, \bibinfo {author} {\bibfnamefont {Z.}~\bibnamefont {Wu}}, \bibinfo {author} {\bibfnamefont {M.}~\bibnamefont {Isaza-Monsalve}}, \bibinfo {author} {\bibfnamefont {S.}~\bibnamefont {Walser}}, \bibinfo {author} {\bibfnamefont {E.}~\bibnamefont {Mattivi}}, \bibinfo {author} {\bibfnamefont {R.}~\bibnamefont {Nardi}},\ and\ \bibinfo {author} {\bibfnamefont {P.}~\bibnamefont {Schindler}},\ }\bibfield  {title} {\bibinfo {title} {Strategies for implementing quantum error correction in molecular rotation},\ }\href@noop {} {\bibfield  {journal} {\bibinfo  {journal} {Quantum}\ }\textbf {\bibinfo {volume} {8}},\ \bibinfo {pages} {1578} (\bibinfo {year} {2024})}\BibitemShut {NoStop}%
\bibitem [{\citenamefont {DeBry}\ \emph {et~al.}(2026)\citenamefont {DeBry}, \citenamefont {Meister}, \citenamefont {Valdes~Martinez}, \citenamefont {Bruzewicz}, \citenamefont {Shi}, \citenamefont {Reens}, \citenamefont {McConnell}, \citenamefont {Chuang},\ and\ \citenamefont {Chiaverini}}]{DeBry2026}%
  \BibitemOpen
  \bibfield  {author} {\bibinfo {author} {\bibfnamefont {K.}~\bibnamefont {DeBry}}, \bibinfo {author} {\bibfnamefont {N.}~\bibnamefont {Meister}}, \bibinfo {author} {\bibfnamefont {A.}~\bibnamefont {Valdes~Martinez}}, \bibinfo {author} {\bibfnamefont {C.~D.}\ \bibnamefont {Bruzewicz}}, \bibinfo {author} {\bibfnamefont {X.}~\bibnamefont {Shi}}, \bibinfo {author} {\bibfnamefont {D.}~\bibnamefont {Reens}}, \bibinfo {author} {\bibfnamefont {R.}~\bibnamefont {McConnell}}, \bibinfo {author} {\bibfnamefont {I.~L.}\ \bibnamefont {Chuang}},\ and\ \bibinfo {author} {\bibfnamefont {J.}~\bibnamefont {Chiaverini}},\ }\bibfield  {title} {\bibinfo {title} {Error correction of a logical qubit encoded in a single atomic ion},\ }\href@noop {} {\bibfield  {journal} {\bibinfo  {journal} {Nature Physics}\ } (\bibinfo {year} {2026})}\BibitemShut {NoStop}%
\bibitem [{\citenamefont {Li}\ \emph {et~al.}(2026)\citenamefont {Li}, \citenamefont {Mei}, \citenamefont {Jie}, \citenamefont {Cai}, \citenamefont {Li}, \citenamefont {Liu}, \citenamefont {Chen}, \citenamefont {Xie}, \citenamefont {Cheng}, \citenamefont {Zhao}, \citenamefont {Luo}, \citenamefont {Zhang}, \citenamefont {Zou}, \citenamefont {Zou}, \citenamefont {Lin},\ and\ \citenamefont {Du}}]{Li2026}%
  \BibitemOpen
  \bibfield  {author} {\bibinfo {author} {\bibfnamefont {Y.}~\bibnamefont {Li}}, \bibinfo {author} {\bibfnamefont {Q.}~\bibnamefont {Mei}}, \bibinfo {author} {\bibfnamefont {Q.-X.}\ \bibnamefont {Jie}}, \bibinfo {author} {\bibfnamefont {W.}~\bibnamefont {Cai}}, \bibinfo {author} {\bibfnamefont {Y.}~\bibnamefont {Li}}, \bibinfo {author} {\bibfnamefont {Z.}~\bibnamefont {Liu}}, \bibinfo {author} {\bibfnamefont {Z.-J.}\ \bibnamefont {Chen}}, \bibinfo {author} {\bibfnamefont {Z.}~\bibnamefont {Xie}}, \bibinfo {author} {\bibfnamefont {X.}~\bibnamefont {Cheng}}, \bibinfo {author} {\bibfnamefont {X.}~\bibnamefont {Zhao}}, \bibinfo {author} {\bibfnamefont {Z.}~\bibnamefont {Luo}}, \bibinfo {author} {\bibfnamefont {M.}~\bibnamefont {Zhang}}, \bibinfo {author} {\bibfnamefont {X.-B.}\ \bibnamefont {Zou}}, \bibinfo {author} {\bibfnamefont {C.-L.}\ \bibnamefont {Zou}}, \bibinfo {author} {\bibfnamefont {Y.}~\bibnamefont {Lin}},\ and\ \bibinfo {author} {\bibfnamefont {J.}~\bibnamefont {Du}},\ }\bibfield  {title} {\bibinfo
  {title} {Beating the break-even point with autonomous quantum error correction},\ }\href@noop {} {\bibfield  {journal} {\bibinfo  {journal} {Nature Physics}\ } (\bibinfo {year} {2026})}\BibitemShut {NoStop}%
\bibitem [{\citenamefont {Brennen}\ \emph {et~al.}(2005)\citenamefont {Brennen}, \citenamefont {O'Leary},\ and\ \citenamefont {Bullock}}]{Brennen2005}%
  \BibitemOpen
  \bibfield  {author} {\bibinfo {author} {\bibfnamefont {G.~K.}\ \bibnamefont {Brennen}}, \bibinfo {author} {\bibfnamefont {D.~P.}\ \bibnamefont {O'Leary}},\ and\ \bibinfo {author} {\bibfnamefont {S.~S.}\ \bibnamefont {Bullock}},\ }\bibfield  {title} {\bibinfo {title} {Criteria for exact qudit universality},\ }\href@noop {} {\bibfield  {journal} {\bibinfo  {journal} {Physical Review A}\ }\textbf {\bibinfo {volume} {71}},\ \bibinfo {pages} {052318} (\bibinfo {year} {2005})}\BibitemShut {NoStop}%
\bibitem [{\citenamefont {O'Leary}\ \emph {et~al.}(2006)\citenamefont {O'Leary}, \citenamefont {Brennen},\ and\ \citenamefont {Bullock}}]{OLeary2006}%
  \BibitemOpen
  \bibfield  {author} {\bibinfo {author} {\bibfnamefont {D.~P.}\ \bibnamefont {O'Leary}}, \bibinfo {author} {\bibfnamefont {G.~K.}\ \bibnamefont {Brennen}},\ and\ \bibinfo {author} {\bibfnamefont {S.~S.}\ \bibnamefont {Bullock}},\ }\bibfield  {title} {\bibinfo {title} {Parallelism for quantum computation with qudits},\ }\href@noop {} {\bibfield  {journal} {\bibinfo  {journal} {Physical Review A}\ }\textbf {\bibinfo {volume} {74}},\ \bibinfo {pages} {032334} (\bibinfo {year} {2006})}\BibitemShut {NoStop}%
\bibitem [{\citenamefont {Neeley}\ \emph {et~al.}(2009)\citenamefont {Neeley}, \citenamefont {Ansmann}, \citenamefont {Bialczak}, \citenamefont {Hofheinz}, \citenamefont {Lucero}, \citenamefont {O'Connell}, \citenamefont {Sank}, \citenamefont {Wang}, \citenamefont {Wenner}, \citenamefont {Cleland}, \citenamefont {Geller},\ and\ \citenamefont {Martinis}}]{Neeley2009}%
  \BibitemOpen
  \bibfield  {author} {\bibinfo {author} {\bibfnamefont {M.}~\bibnamefont {Neeley}}, \bibinfo {author} {\bibfnamefont {M.}~\bibnamefont {Ansmann}}, \bibinfo {author} {\bibfnamefont {R.~C.}\ \bibnamefont {Bialczak}}, \bibinfo {author} {\bibfnamefont {M.}~\bibnamefont {Hofheinz}}, \bibinfo {author} {\bibfnamefont {E.}~\bibnamefont {Lucero}}, \bibinfo {author} {\bibfnamefont {A.~D.}\ \bibnamefont {O'Connell}}, \bibinfo {author} {\bibfnamefont {D.}~\bibnamefont {Sank}}, \bibinfo {author} {\bibfnamefont {H.}~\bibnamefont {Wang}}, \bibinfo {author} {\bibfnamefont {J.}~\bibnamefont {Wenner}}, \bibinfo {author} {\bibfnamefont {A.~N.}\ \bibnamefont {Cleland}}, \bibinfo {author} {\bibfnamefont {M.~R.}\ \bibnamefont {Geller}},\ and\ \bibinfo {author} {\bibfnamefont {J.~M.}\ \bibnamefont {Martinis}},\ }\bibfield  {title} {\bibinfo {title} {Emulation of a quantum spin with a superconducting phase qudit},\ }\href@noop {} {\bibfield  {journal} {\bibinfo  {journal} {Science}\ }\textbf {\bibinfo {volume} {325}},\ \bibinfo
  {pages} {722} (\bibinfo {year} {2009})}\BibitemShut {NoStop}%
\bibitem [{\citenamefont {Blok}\ \emph {et~al.}(2021)\citenamefont {Blok}, \citenamefont {Ramasesh}, \citenamefont {Schuster}, \citenamefont {O'Brien}, \citenamefont {Kreikebaum}, \citenamefont {Dahlen}, \citenamefont {Morvan}, \citenamefont {Yoshida}, \citenamefont {Yao},\ and\ \citenamefont {Siddiqi}}]{Blok2021}%
  \BibitemOpen
  \bibfield  {author} {\bibinfo {author} {\bibfnamefont {M.}~\bibnamefont {Blok}}, \bibinfo {author} {\bibfnamefont {V.}~\bibnamefont {Ramasesh}}, \bibinfo {author} {\bibfnamefont {T.}~\bibnamefont {Schuster}}, \bibinfo {author} {\bibfnamefont {K.}~\bibnamefont {O'Brien}}, \bibinfo {author} {\bibfnamefont {J.}~\bibnamefont {Kreikebaum}}, \bibinfo {author} {\bibfnamefont {D.}~\bibnamefont {Dahlen}}, \bibinfo {author} {\bibfnamefont {A.}~\bibnamefont {Morvan}}, \bibinfo {author} {\bibfnamefont {B.}~\bibnamefont {Yoshida}}, \bibinfo {author} {\bibfnamefont {N.}~\bibnamefont {Yao}},\ and\ \bibinfo {author} {\bibfnamefont {I.}~\bibnamefont {Siddiqi}},\ }\bibfield  {title} {\bibinfo {title} {Quantum information scrambling on a superconducting qutrit processor},\ }\href@noop {} {\bibfield  {journal} {\bibinfo  {journal} {Physical Review X}\ }\textbf {\bibinfo {volume} {11}},\ \bibinfo {pages} {021010} (\bibinfo {year} {2021})}\BibitemShut {NoStop}%
\bibitem [{\citenamefont {Barnes}\ \emph {et~al.}(2022)\citenamefont {Barnes}, \citenamefont {Battaglino}, \citenamefont {Bloom}, \citenamefont {Cassella}, \citenamefont {Coxe}, \citenamefont {Crisosto}, \citenamefont {King}, \citenamefont {Kondov}, \citenamefont {Kotru}, \citenamefont {Larsen}, \citenamefont {Lauigan}, \citenamefont {Lester}, \citenamefont {McDonald}, \citenamefont {Megidish}, \citenamefont {Narayanaswami}, \citenamefont {Nishiguchi}, \citenamefont {Notermans}, \citenamefont {Peng}, \citenamefont {Ryou}, \citenamefont {Wu},\ and\ \citenamefont {Yarwood}}]{Barnes2022}%
  \BibitemOpen
  \bibfield  {author} {\bibinfo {author} {\bibfnamefont {K.}~\bibnamefont {Barnes}}, \bibinfo {author} {\bibfnamefont {P.}~\bibnamefont {Battaglino}}, \bibinfo {author} {\bibfnamefont {B.~J.}\ \bibnamefont {Bloom}}, \bibinfo {author} {\bibfnamefont {K.}~\bibnamefont {Cassella}}, \bibinfo {author} {\bibfnamefont {R.}~\bibnamefont {Coxe}}, \bibinfo {author} {\bibfnamefont {N.}~\bibnamefont {Crisosto}}, \bibinfo {author} {\bibfnamefont {J.~P.}\ \bibnamefont {King}}, \bibinfo {author} {\bibfnamefont {S.~S.}\ \bibnamefont {Kondov}}, \bibinfo {author} {\bibfnamefont {K.}~\bibnamefont {Kotru}}, \bibinfo {author} {\bibfnamefont {S.~C.}\ \bibnamefont {Larsen}}, \bibinfo {author} {\bibfnamefont {J.}~\bibnamefont {Lauigan}}, \bibinfo {author} {\bibfnamefont {B.~J.}\ \bibnamefont {Lester}}, \bibinfo {author} {\bibfnamefont {M.}~\bibnamefont {McDonald}}, \bibinfo {author} {\bibfnamefont {E.}~\bibnamefont {Megidish}}, \bibinfo {author} {\bibfnamefont {S.}~\bibnamefont {Narayanaswami}}, \bibinfo {author} {\bibfnamefont
  {C.}~\bibnamefont {Nishiguchi}}, \bibinfo {author} {\bibfnamefont {R.}~\bibnamefont {Notermans}}, \bibinfo {author} {\bibfnamefont {L.~S.}\ \bibnamefont {Peng}}, \bibinfo {author} {\bibfnamefont {A.}~\bibnamefont {Ryou}}, \bibinfo {author} {\bibfnamefont {T.-Y.}\ \bibnamefont {Wu}},\ and\ \bibinfo {author} {\bibfnamefont {M.}~\bibnamefont {Yarwood}},\ }\bibfield  {title} {\bibinfo {title} {Assembly and coherent control of a register of nuclear spin qubits},\ }\href@noop {} {\bibfield  {journal} {\bibinfo  {journal} {Nature Communications}\ }\textbf {\bibinfo {volume} {13}},\ \bibinfo {pages} {2779} (\bibinfo {year} {2022})}\BibitemShut {NoStop}%
\bibitem [{\citenamefont {Krondorfer}\ \emph {et~al.}(2026)\citenamefont {Krondorfer}, \citenamefont {Diez}, \citenamefont {Kruckenhauser},\ and\ \citenamefont {Hauser}}]{Krondorfer2026}%
  \BibitemOpen
  \bibfield  {author} {\bibinfo {author} {\bibfnamefont {J.~K.}\ \bibnamefont {Krondorfer}}, \bibinfo {author} {\bibfnamefont {M.}~\bibnamefont {Diez}}, \bibinfo {author} {\bibfnamefont {A.}~\bibnamefont {Kruckenhauser}},\ and\ \bibinfo {author} {\bibfnamefont {A.~W.}\ \bibnamefont {Hauser}},\ }\bibfield  {title} {\bibinfo {title} {All-optical universal control of nuclear-spin qudits in trapped neutral atoms},\ }\href@noop {} {\bibfield  {journal} {\bibinfo  {journal} {Quantum Science and Technology}\ }\textbf {\bibinfo {volume} {11}},\ \bibinfo {pages} {045042} (\bibinfo {year} {2026})}\BibitemShut {NoStop}%
\bibitem [{\citenamefont {Ahmed}\ \emph {et~al.}(2025)\citenamefont {Ahmed}, \citenamefont {Litvinov}, \citenamefont {Guesdon}, \citenamefont {Mar{\'e}chal}, \citenamefont {Huckans}, \citenamefont {Pasquiou}, \citenamefont {Laburthe-Tolra},\ and\ \citenamefont {Robert-de Saint-Vincent}}]{Ahmed2025}%
  \BibitemOpen
  \bibfield  {author} {\bibinfo {author} {\bibfnamefont {H.}~\bibnamefont {Ahmed}}, \bibinfo {author} {\bibfnamefont {A.}~\bibnamefont {Litvinov}}, \bibinfo {author} {\bibfnamefont {P.}~\bibnamefont {Guesdon}}, \bibinfo {author} {\bibfnamefont {E.}~\bibnamefont {Mar{\'e}chal}}, \bibinfo {author} {\bibfnamefont {J.}~\bibnamefont {Huckans}}, \bibinfo {author} {\bibfnamefont {B.}~\bibnamefont {Pasquiou}}, \bibinfo {author} {\bibfnamefont {B.}~\bibnamefont {Laburthe-Tolra}},\ and\ \bibinfo {author} {\bibfnamefont {M.}~\bibnamefont {Robert-de Saint-Vincent}},\ }\bibfield  {title} {\bibinfo {title} {Coherent control over the high-dimensional space of the nuclear spin of alkaline-earth atoms},\ }\href@noop {} {\bibfield  {journal} {\bibinfo  {journal} {PRX Quantum}\ }\textbf {\bibinfo {volume} {6}},\ \bibinfo {pages} {020352} (\bibinfo {year} {2025})}\BibitemShut {NoStop}%
\bibitem [{\citenamefont {Krondorfer}\ \emph {et~al.}(2025{\natexlab{a}})\citenamefont {Krondorfer}, \citenamefont {Pucher}, \citenamefont {Diez}, \citenamefont {Blatt},\ and\ \citenamefont {Hauser}}]{Krondorfer2025a}%
  \BibitemOpen
  \bibfield  {author} {\bibinfo {author} {\bibfnamefont {J.~K.}\ \bibnamefont {Krondorfer}}, \bibinfo {author} {\bibfnamefont {S.}~\bibnamefont {Pucher}}, \bibinfo {author} {\bibfnamefont {M.}~\bibnamefont {Diez}}, \bibinfo {author} {\bibfnamefont {S.}~\bibnamefont {Blatt}},\ and\ \bibinfo {author} {\bibfnamefont {A.~W.}\ \bibnamefont {Hauser}},\ }\bibfield  {title} {\bibinfo {title} {Optical nuclear electric resonance as single qubit gate for trapped neutral atoms},\ }\href@noop {} {\bibfield  {journal} {\bibinfo  {journal} {Journal of Physics B: Atomic, Molecular and Optical Physics}\ }\textbf {\bibinfo {volume} {58}},\ \bibinfo {pages} {235001} (\bibinfo {year} {2025}{\natexlab{a}})}\BibitemShut {NoStop}%
\bibitem [{\citenamefont {Krondorfer}\ \emph {et~al.}(2025{\natexlab{b}})\citenamefont {Krondorfer}, \citenamefont {Diez},\ and\ \citenamefont {Hauser}}]{Krondorfer2025b}%
  \BibitemOpen
  \bibfield  {author} {\bibinfo {author} {\bibfnamefont {J.~K.}\ \bibnamefont {Krondorfer}}, \bibinfo {author} {\bibfnamefont {M.}~\bibnamefont {Diez}},\ and\ \bibinfo {author} {\bibfnamefont {A.~W.}\ \bibnamefont {Hauser}},\ }\bibfield  {title} {\bibinfo {title} {Single-qudit control in {$^{87}$Sr} via optical nuclear electric resonance},\ }\href@noop {} {\bibfield  {journal} {\bibinfo  {journal} {Physical Review A}\ }\textbf {\bibinfo {volume} {112}},\ \bibinfo {pages} {062614} (\bibinfo {year} {2025}{\natexlab{b}})}\BibitemShut {NoStop}%
\bibitem [{\citenamefont {Heizenreder}\ \emph {et~al.}(2026)\citenamefont {Heizenreder}, \citenamefont {Gerritsen}, \citenamefont {Fouka}, \citenamefont {Spreeuw}, \citenamefont {Schreck}, \citenamefont {Naini},\ and\ \citenamefont {Urech}}]{Heizenreder2026}%
  \BibitemOpen
  \bibfield  {author} {\bibinfo {author} {\bibfnamefont {B.}~\bibnamefont {Heizenreder}}, \bibinfo {author} {\bibfnamefont {B.}~\bibnamefont {Gerritsen}}, \bibinfo {author} {\bibfnamefont {K.}~\bibnamefont {Fouka}}, \bibinfo {author} {\bibfnamefont {R.~J.~C.}\ \bibnamefont {Spreeuw}}, \bibinfo {author} {\bibfnamefont {F.}~\bibnamefont {Schreck}}, \bibinfo {author} {\bibfnamefont {A.~S.}\ \bibnamefont {Naini}},\ and\ \bibinfo {author} {\bibfnamefont {A.}~\bibnamefont {Urech}},\ }\bibfield  {title} {\bibinfo {title} {Engineering {Z}eeman-manifold quintets using state-dependent light shifts in neutral atoms},\ }\href@noop {} {\bibfield  {journal} {\bibinfo  {journal} {Physical Review A}\ }\textbf {\bibinfo {volume} {114}},\ \bibinfo {pages} {042614} (\bibinfo {year} {2026})}\BibitemShut {NoStop}%
\bibitem [{\citenamefont {Castro}\ \emph {et~al.}(2022)\citenamefont {Castro}, \citenamefont {Garc{\'\i}a~Carrizo}, \citenamefont {Roca}, \citenamefont {Zueco},\ and\ \citenamefont {Luis}}]{Castro2022}%
  \BibitemOpen
  \bibfield  {author} {\bibinfo {author} {\bibfnamefont {A.}~\bibnamefont {Castro}}, \bibinfo {author} {\bibfnamefont {A.}~\bibnamefont {Garc{\'\i}a~Carrizo}}, \bibinfo {author} {\bibfnamefont {S.}~\bibnamefont {Roca}}, \bibinfo {author} {\bibfnamefont {D.}~\bibnamefont {Zueco}},\ and\ \bibinfo {author} {\bibfnamefont {F.}~\bibnamefont {Luis}},\ }\bibfield  {title} {\bibinfo {title} {Optimal control of molecular spin qudits},\ }\href@noop {} {\bibfield  {journal} {\bibinfo  {journal} {Physical Review Applied}\ }\textbf {\bibinfo {volume} {17}},\ \bibinfo {pages} {064028} (\bibinfo {year} {2022})}\BibitemShut {NoStop}%
\bibitem [{\citenamefont {Ringbauer}\ \emph {et~al.}(2022)\citenamefont {Ringbauer}, \citenamefont {Meth}, \citenamefont {Postler}, \citenamefont {Stricker}, \citenamefont {Blatt}, \citenamefont {Schindler},\ and\ \citenamefont {Monz}}]{Ringbauer2022}%
  \BibitemOpen
  \bibfield  {author} {\bibinfo {author} {\bibfnamefont {M.}~\bibnamefont {Ringbauer}}, \bibinfo {author} {\bibfnamefont {M.}~\bibnamefont {Meth}}, \bibinfo {author} {\bibfnamefont {L.}~\bibnamefont {Postler}}, \bibinfo {author} {\bibfnamefont {R.}~\bibnamefont {Stricker}}, \bibinfo {author} {\bibfnamefont {R.}~\bibnamefont {Blatt}}, \bibinfo {author} {\bibfnamefont {P.}~\bibnamefont {Schindler}},\ and\ \bibinfo {author} {\bibfnamefont {T.}~\bibnamefont {Monz}},\ }\bibfield  {title} {\bibinfo {title} {A universal qudit quantum processor with trapped ions},\ }\href@noop {} {\bibfield  {journal} {\bibinfo  {journal} {Nature Physics}\ }\textbf {\bibinfo {volume} {18}},\ \bibinfo {pages} {1053} (\bibinfo {year} {2022})}\BibitemShut {NoStop}%
\bibitem [{\citenamefont {Hrmo}\ \emph {et~al.}(2023)\citenamefont {Hrmo}, \citenamefont {Wilhelm}, \citenamefont {Gerster}, \citenamefont {van Mourik}, \citenamefont {Huber}, \citenamefont {Blatt}, \citenamefont {Schindler}, \citenamefont {Monz},\ and\ \citenamefont {Ringbauer}}]{Hrmo2023}%
  \BibitemOpen
  \bibfield  {author} {\bibinfo {author} {\bibfnamefont {P.}~\bibnamefont {Hrmo}}, \bibinfo {author} {\bibfnamefont {B.}~\bibnamefont {Wilhelm}}, \bibinfo {author} {\bibfnamefont {L.}~\bibnamefont {Gerster}}, \bibinfo {author} {\bibfnamefont {M.~W.}\ \bibnamefont {van Mourik}}, \bibinfo {author} {\bibfnamefont {M.}~\bibnamefont {Huber}}, \bibinfo {author} {\bibfnamefont {R.}~\bibnamefont {Blatt}}, \bibinfo {author} {\bibfnamefont {P.}~\bibnamefont {Schindler}}, \bibinfo {author} {\bibfnamefont {T.}~\bibnamefont {Monz}},\ and\ \bibinfo {author} {\bibfnamefont {M.}~\bibnamefont {Ringbauer}},\ }\bibfield  {title} {\bibinfo {title} {Native qudit entanglement in a trapped ion quantum processor},\ }\href@noop {} {\bibfield  {journal} {\bibinfo  {journal} {Nature Communications}\ }\textbf {\bibinfo {volume} {14}},\ \bibinfo {pages} {2242} (\bibinfo {year} {2023})}\BibitemShut {NoStop}%
\bibitem [{\citenamefont {Low}\ \emph {et~al.}(2025)\citenamefont {Low}, \citenamefont {White},\ and\ \citenamefont {Senko}}]{Low2025}%
  \BibitemOpen
  \bibfield  {author} {\bibinfo {author} {\bibfnamefont {P.~J.}\ \bibnamefont {Low}}, \bibinfo {author} {\bibfnamefont {B.}~\bibnamefont {White}},\ and\ \bibinfo {author} {\bibfnamefont {C.}~\bibnamefont {Senko}},\ }\bibfield  {title} {\bibinfo {title} {Control and readout of a 13-level trapped ion qudit},\ }\href@noop {} {\bibfield  {journal} {\bibinfo  {journal} {npj Quantum Information}\ }\textbf {\bibinfo {volume} {11}},\ \bibinfo {pages} {85} (\bibinfo {year} {2025})}\BibitemShut {NoStop}%
\bibitem [{\citenamefont {Low}\ \emph {et~al.}(2026)\citenamefont {Low}, \citenamefont {Zutt}, \citenamefont {Tathed},\ and\ \citenamefont {Senko}}]{Low2026}%
  \BibitemOpen
  \bibfield  {author} {\bibinfo {author} {\bibfnamefont {P.~J.}\ \bibnamefont {Low}}, \bibinfo {author} {\bibfnamefont {N.~C.~F.}\ \bibnamefont {Zutt}}, \bibinfo {author} {\bibfnamefont {G.~A.}\ \bibnamefont {Tathed}},\ and\ \bibinfo {author} {\bibfnamefont {C.}~\bibnamefont {Senko}},\ }\bibfield  {title} {\bibinfo {title} {Quantum logic operations and algorithms in a single 25-level atomic qudit},\ }\href@noop {} {\bibfield  {journal} {\bibinfo  {journal} {Nature Communications}\ }\textbf {\bibinfo {volume} {17}},\ \bibinfo {pages} {7098} (\bibinfo {year} {2026})}\BibitemShut {NoStop}%
\bibitem [{\citenamefont {Burshtein}\ \emph {et~al.}(2026)\citenamefont {Burshtein}, \citenamefont {Fraenkel}, \citenamefont {Goldstein},\ and\ \citenamefont {Finkelstein}}]{Burshtein2026}%
  \BibitemOpen
  \bibfield  {author} {\bibinfo {author} {\bibfnamefont {A.}~\bibnamefont {Burshtein}}, \bibinfo {author} {\bibfnamefont {S.}~\bibnamefont {Fraenkel}}, \bibinfo {author} {\bibfnamefont {M.}~\bibnamefont {Goldstein}},\ and\ \bibinfo {author} {\bibfnamefont {J.}~\bibnamefont {Finkelstein}},\ }\bibfield  {title} {\bibinfo {title} {Robust control and entanglement of qudits in neutral atom arrays},\ }\href@noop {} {\bibfield  {journal} {\bibinfo  {journal} {Physical Review Research}\ }\textbf {\bibinfo {volume} {8}},\ \bibinfo {pages} {013055} (\bibinfo {year} {2026})}\BibitemShut {NoStop}%
\bibitem [{\citenamefont {Gregory}\ \emph {et~al.}(2026)\citenamefont {Gregory}, \citenamefont {Ritchie}, \citenamefont {Quinn}, \citenamefont {Brudney}, \citenamefont {Wineland}, \citenamefont {Allcock},\ and\ \citenamefont {O'Reilly}}]{Gregory2026}%
  \BibitemOpen
  \bibfield  {author} {\bibinfo {author} {\bibfnamefont {G.~J.}\ \bibnamefont {Gregory}}, \bibinfo {author} {\bibfnamefont {E.~R.}\ \bibnamefont {Ritchie}}, \bibinfo {author} {\bibfnamefont {A.}~\bibnamefont {Quinn}}, \bibinfo {author} {\bibfnamefont {S.}~\bibnamefont {Brudney}}, \bibinfo {author} {\bibfnamefont {D.~J.}\ \bibnamefont {Wineland}}, \bibinfo {author} {\bibfnamefont {D.~T.~C.}\ \bibnamefont {Allcock}},\ and\ \bibinfo {author} {\bibfnamefont {J.}~\bibnamefont {O'Reilly}},\ }\href@noop {} {\bibinfo {title} {Four- and six-photon stimulated {R}aman transitions for coherent qubit and qudit operations}} (\bibinfo {year} {2026}),\ \Eprint {https://arxiv.org/abs/2602.18567} {arXiv:2602.18567 [quant-ph]} \BibitemShut {NoStop}%
\bibitem [{\citenamefont {Shi}\ \emph {et~al.}(2026)\citenamefont {Shi}, \citenamefont {Sinanan-Singh}, \citenamefont {Burke}, \citenamefont {Chiaverini},\ and\ \citenamefont {Chuang}}]{Shi2026}%
  \BibitemOpen
  \bibfield  {author} {\bibinfo {author} {\bibfnamefont {X.}~\bibnamefont {Shi}}, \bibinfo {author} {\bibfnamefont {J.}~\bibnamefont {Sinanan-Singh}}, \bibinfo {author} {\bibfnamefont {T.~J.}\ \bibnamefont {Burke}}, \bibinfo {author} {\bibfnamefont {J.}~\bibnamefont {Chiaverini}},\ and\ \bibinfo {author} {\bibfnamefont {I.~L.}\ \bibnamefont {Chuang}},\ }\bibfield  {title} {\bibinfo {title} {Efficient implementation of a quantum algorithm with a trapped ion qudit},\ }\href@noop {} {\bibfield  {journal} {\bibinfo  {journal} {Nature Communications}\ }\textbf {\bibinfo {volume} {17}},\ \bibinfo {pages} {1911} (\bibinfo {year} {2026})}\BibitemShut {NoStop}%
\bibitem [{\citenamefont {Champion}\ \emph {et~al.}(2025{\natexlab{b}})\citenamefont {Champion}, \citenamefont {Wang}, \citenamefont {Parker},\ and\ \citenamefont {Blok}}]{Champion2025}%
  \BibitemOpen
  \bibfield  {author} {\bibinfo {author} {\bibfnamefont {E.}~\bibnamefont {Champion}}, \bibinfo {author} {\bibfnamefont {Z.}~\bibnamefont {Wang}}, \bibinfo {author} {\bibfnamefont {R.~W.}\ \bibnamefont {Parker}},\ and\ \bibinfo {author} {\bibfnamefont {M.~S.}\ \bibnamefont {Blok}},\ }\bibfield  {title} {\bibinfo {title} {Efficient control of a transmon qudit using effective spin-{$7/2$} rotations},\ }\href@noop {} {\bibfield  {journal} {\bibinfo  {journal} {Physical Review X}\ }\textbf {\bibinfo {volume} {15}},\ \bibinfo {pages} {021096} (\bibinfo {year} {2025}{\natexlab{b}})}\BibitemShut {NoStop}%
\bibitem [{\citenamefont {Yu}\ \emph {et~al.}(2025)\citenamefont {Yu}, \citenamefont {Wilhelm}, \citenamefont {Holmes}, \citenamefont {Vaartjes}, \citenamefont {Schwienbacher}, \citenamefont {Nurizzo}, \citenamefont {Kringh{\o}j}, \citenamefont {van Blankenstein}, \citenamefont {Jakob}, \citenamefont {Gupta}, \citenamefont {Hudson}, \citenamefont {Itoh}, \citenamefont {Murray}, \citenamefont {Blume-Kohout}, \citenamefont {Ladd}, \citenamefont {Anand}, \citenamefont {Dzurak}, \citenamefont {Sanders}, \citenamefont {Jamieson},\ and\ \citenamefont {Morello}}]{Yu2025}%
  \BibitemOpen
  \bibfield  {author} {\bibinfo {author} {\bibfnamefont {X.}~\bibnamefont {Yu}}, \bibinfo {author} {\bibfnamefont {B.}~\bibnamefont {Wilhelm}}, \bibinfo {author} {\bibfnamefont {D.}~\bibnamefont {Holmes}}, \bibinfo {author} {\bibfnamefont {A.}~\bibnamefont {Vaartjes}}, \bibinfo {author} {\bibfnamefont {D.}~\bibnamefont {Schwienbacher}}, \bibinfo {author} {\bibfnamefont {M.}~\bibnamefont {Nurizzo}}, \bibinfo {author} {\bibfnamefont {A.}~\bibnamefont {Kringh{\o}j}}, \bibinfo {author} {\bibfnamefont {M.~R.}\ \bibnamefont {van Blankenstein}}, \bibinfo {author} {\bibfnamefont {A.~M.}\ \bibnamefont {Jakob}}, \bibinfo {author} {\bibfnamefont {P.}~\bibnamefont {Gupta}}, \bibinfo {author} {\bibfnamefont {F.~E.}\ \bibnamefont {Hudson}}, \bibinfo {author} {\bibfnamefont {K.~M.}\ \bibnamefont {Itoh}}, \bibinfo {author} {\bibfnamefont {R.~J.}\ \bibnamefont {Murray}}, \bibinfo {author} {\bibfnamefont {R.}~\bibnamefont {Blume-Kohout}}, \bibinfo {author} {\bibfnamefont {T.~D.}\ \bibnamefont {Ladd}}, \bibinfo {author}
  {\bibfnamefont {N.}~\bibnamefont {Anand}}, \bibinfo {author} {\bibfnamefont {A.~S.}\ \bibnamefont {Dzurak}}, \bibinfo {author} {\bibfnamefont {B.~C.}\ \bibnamefont {Sanders}}, \bibinfo {author} {\bibfnamefont {D.~N.}\ \bibnamefont {Jamieson}},\ and\ \bibinfo {author} {\bibfnamefont {A.}~\bibnamefont {Morello}},\ }\bibfield  {title} {\bibinfo {title} {Schr{\"o}dinger cat states of a nuclear spin qudit in silicon},\ }\href@noop {} {\bibfield  {journal} {\bibinfo  {journal} {Nature Physics}\ } (\bibinfo {year} {2025})}\BibitemShut {NoStop}%
\bibitem [{\citenamefont {Lim}\ \emph {et~al.}(2025)\citenamefont {Lim}, \citenamefont {Vaganov}, \citenamefont {Liu},\ and\ \citenamefont {Ardavan}}]{Lim2025}%
  \BibitemOpen
  \bibfield  {author} {\bibinfo {author} {\bibfnamefont {S.}~\bibnamefont {Lim}}, \bibinfo {author} {\bibfnamefont {M.~V.}\ \bibnamefont {Vaganov}}, \bibinfo {author} {\bibfnamefont {J.}~\bibnamefont {Liu}},\ and\ \bibinfo {author} {\bibfnamefont {A.}~\bibnamefont {Ardavan}},\ }\bibfield  {title} {\bibinfo {title} {Demonstrating experimentally the encoding and dynamics of an error-correctable logical qubit on a hyperfine-coupled nuclear spin qudit},\ }\href@noop {} {\bibfield  {journal} {\bibinfo  {journal} {Physical Review Letters}\ }\textbf {\bibinfo {volume} {134}},\ \bibinfo {pages} {070603} (\bibinfo {year} {2025})}\BibitemShut {NoStop}%
\bibitem [{\citenamefont {Fern{\'a}ndez~de Fuentes}\ \emph {et~al.}(2024)\citenamefont {Fern{\'a}ndez~de Fuentes}, \citenamefont {Botzem}, \citenamefont {Johnson}, \citenamefont {Vaartjes}, \citenamefont {Asaad}, \citenamefont {Mourik}, \citenamefont {Hudson}, \citenamefont {Itoh}, \citenamefont {Johnson}, \citenamefont {Jakob}, \citenamefont {McCallum}, \citenamefont {Jamieson}, \citenamefont {Dzurak},\ and\ \citenamefont {Morello}}]{FernandezDeFuentes2024}%
  \BibitemOpen
  \bibfield  {author} {\bibinfo {author} {\bibfnamefont {I.}~\bibnamefont {Fern{\'a}ndez~de Fuentes}}, \bibinfo {author} {\bibfnamefont {T.}~\bibnamefont {Botzem}}, \bibinfo {author} {\bibfnamefont {M.~A.~I.}\ \bibnamefont {Johnson}}, \bibinfo {author} {\bibfnamefont {A.}~\bibnamefont {Vaartjes}}, \bibinfo {author} {\bibfnamefont {S.}~\bibnamefont {Asaad}}, \bibinfo {author} {\bibfnamefont {V.}~\bibnamefont {Mourik}}, \bibinfo {author} {\bibfnamefont {F.~E.}\ \bibnamefont {Hudson}}, \bibinfo {author} {\bibfnamefont {K.~M.}\ \bibnamefont {Itoh}}, \bibinfo {author} {\bibfnamefont {B.~C.}\ \bibnamefont {Johnson}}, \bibinfo {author} {\bibfnamefont {A.~M.}\ \bibnamefont {Jakob}}, \bibinfo {author} {\bibfnamefont {J.~C.}\ \bibnamefont {McCallum}}, \bibinfo {author} {\bibfnamefont {D.~N.}\ \bibnamefont {Jamieson}}, \bibinfo {author} {\bibfnamefont {A.~S.}\ \bibnamefont {Dzurak}},\ and\ \bibinfo {author} {\bibfnamefont {A.}~\bibnamefont {Morello}},\ }\bibfield  {title} {\bibinfo {title} {Navigating the 16-dimensional
  {H}ilbert space of a high-spin donor qudit with electric and magnetic fields},\ }\href@noop {} {\bibfield  {journal} {\bibinfo  {journal} {Nature Communications}\ }\textbf {\bibinfo {volume} {15}},\ \bibinfo {pages} {1380} (\bibinfo {year} {2024})}\BibitemShut {NoStop}%
\bibitem [{\citenamefont {Wang}\ \emph {et~al.}(2025)\citenamefont {Wang}, \citenamefont {Parker}, \citenamefont {Champion},\ and\ \citenamefont {Blok}}]{Wang2025TransmonQudit}%
  \BibitemOpen
  \bibfield  {author} {\bibinfo {author} {\bibfnamefont {Z.}~\bibnamefont {Wang}}, \bibinfo {author} {\bibfnamefont {R.~W.}\ \bibnamefont {Parker}}, \bibinfo {author} {\bibfnamefont {E.}~\bibnamefont {Champion}},\ and\ \bibinfo {author} {\bibfnamefont {M.~S.}\ \bibnamefont {Blok}},\ }\bibfield  {title} {\bibinfo {title} {High-${E}_{J}/{E}_{C}$ transmon qudits with up to 12 levels},\ }\href@noop {} {\bibfield  {journal} {\bibinfo  {journal} {Physical Review Applied}\ }\textbf {\bibinfo {volume} {23}},\ \bibinfo {pages} {034046} (\bibinfo {year} {2025})}\BibitemShut {NoStop}%
\bibitem [{\citenamefont {Merkel}\ \emph {et~al.}(2008)\citenamefont {Merkel}, \citenamefont {Jessen},\ and\ \citenamefont {Deutsch}}]{Merkel2008}%
  \BibitemOpen
  \bibfield  {author} {\bibinfo {author} {\bibfnamefont {S.~T.}\ \bibnamefont {Merkel}}, \bibinfo {author} {\bibfnamefont {P.~S.}\ \bibnamefont {Jessen}},\ and\ \bibinfo {author} {\bibfnamefont {I.~H.}\ \bibnamefont {Deutsch}},\ }\bibfield  {title} {\bibinfo {title} {Quantum control of the hyperfine-coupled electron and nuclear spins in alkali-metal atoms},\ }\href@noop {} {\bibfield  {journal} {\bibinfo  {journal} {Physical Review A}\ }\textbf {\bibinfo {volume} {78}},\ \bibinfo {pages} {023404} (\bibinfo {year} {2008})}\BibitemShut {NoStop}%
\bibitem [{\citenamefont {Smith}\ \emph {et~al.}(2013)\citenamefont {Smith}, \citenamefont {Anderson}, \citenamefont {Sosa-Martinez}, \citenamefont {Riofr{\'\i}o}, \citenamefont {Deutsch},\ and\ \citenamefont {Jessen}}]{Smith2013}%
  \BibitemOpen
  \bibfield  {author} {\bibinfo {author} {\bibfnamefont {A.}~\bibnamefont {Smith}}, \bibinfo {author} {\bibfnamefont {B.~E.}\ \bibnamefont {Anderson}}, \bibinfo {author} {\bibfnamefont {H.}~\bibnamefont {Sosa-Martinez}}, \bibinfo {author} {\bibfnamefont {C.~A.}\ \bibnamefont {Riofr{\'\i}o}}, \bibinfo {author} {\bibfnamefont {I.~H.}\ \bibnamefont {Deutsch}},\ and\ \bibinfo {author} {\bibfnamefont {P.~S.}\ \bibnamefont {Jessen}},\ }\bibfield  {title} {\bibinfo {title} {Quantum control in the {Cs} {$6S_{1/2}$} ground manifold using radio-frequency and microwave magnetic fields},\ }\href@noop {} {\bibfield  {journal} {\bibinfo  {journal} {Physical Review Letters}\ }\textbf {\bibinfo {volume} {111}},\ \bibinfo {pages} {170502} (\bibinfo {year} {2013})}\BibitemShut {NoStop}%
\bibitem [{\citenamefont {Merkel}\ \emph {et~al.}(2009)\citenamefont {Merkel}, \citenamefont {Brennen}, \citenamefont {Jessen},\ and\ \citenamefont {Deutsch}}]{Merkel2009}%
  \BibitemOpen
  \bibfield  {author} {\bibinfo {author} {\bibfnamefont {S.~T.}\ \bibnamefont {Merkel}}, \bibinfo {author} {\bibfnamefont {G.}~\bibnamefont {Brennen}}, \bibinfo {author} {\bibfnamefont {P.~S.}\ \bibnamefont {Jessen}},\ and\ \bibinfo {author} {\bibfnamefont {I.~H.}\ \bibnamefont {Deutsch}},\ }\bibfield  {title} {\bibinfo {title} {Constructing general unitary maps from state preparations},\ }\href@noop {} {\bibfield  {journal} {\bibinfo  {journal} {Physical Review A}\ }\textbf {\bibinfo {volume} {80}},\ \bibinfo {pages} {023424} (\bibinfo {year} {2009})}\BibitemShut {NoStop}%
\bibitem [{\citenamefont {Anderson}\ \emph {et~al.}(2015)\citenamefont {Anderson}, \citenamefont {Sosa-Martinez}, \citenamefont {Riofr{\'\i}o}, \citenamefont {Deutsch},\ and\ \citenamefont {Jessen}}]{Anderson2015}%
  \BibitemOpen
  \bibfield  {author} {\bibinfo {author} {\bibfnamefont {B.}~\bibnamefont {Anderson}}, \bibinfo {author} {\bibfnamefont {H.}~\bibnamefont {Sosa-Martinez}}, \bibinfo {author} {\bibfnamefont {C.}~\bibnamefont {Riofr{\'\i}o}}, \bibinfo {author} {\bibfnamefont {I.~H.}\ \bibnamefont {Deutsch}},\ and\ \bibinfo {author} {\bibfnamefont {P.~S.}\ \bibnamefont {Jessen}},\ }\bibfield  {title} {\bibinfo {title} {Accurate and robust unitary transformations of a high-dimensional quantum system},\ }\href@noop {} {\bibfield  {journal} {\bibinfo  {journal} {Physical Review Letters}\ }\textbf {\bibinfo {volume} {114}},\ \bibinfo {pages} {240401} (\bibinfo {year} {2015})}\BibitemShut {NoStop}%
\bibitem [{\citenamefont {Mischuck}\ \emph {et~al.}(2012)\citenamefont {Mischuck}, \citenamefont {Merkel},\ and\ \citenamefont {Deutsch}}]{Mischuck2012}%
  \BibitemOpen
  \bibfield  {author} {\bibinfo {author} {\bibfnamefont {B.~E.}\ \bibnamefont {Mischuck}}, \bibinfo {author} {\bibfnamefont {S.~T.}\ \bibnamefont {Merkel}},\ and\ \bibinfo {author} {\bibfnamefont {I.~H.}\ \bibnamefont {Deutsch}},\ }\bibfield  {title} {\bibinfo {title} {Control of inhomogeneous atomic ensembles of hyperfine qudits},\ }\href@noop {} {\bibfield  {journal} {\bibinfo  {journal} {Physical Review A}\ }\textbf {\bibinfo {volume} {85}},\ \bibinfo {pages} {022302} (\bibinfo {year} {2012})}\BibitemShut {NoStop}%
\bibitem [{\citenamefont {Chaudhury}\ \emph {et~al.}(2007)\citenamefont {Chaudhury}, \citenamefont {Merkel}, \citenamefont {Herr}, \citenamefont {Silberfarb}, \citenamefont {Deutsch},\ and\ \citenamefont {Jessen}}]{Chaudhury2007}%
  \BibitemOpen
  \bibfield  {author} {\bibinfo {author} {\bibfnamefont {S.}~\bibnamefont {Chaudhury}}, \bibinfo {author} {\bibfnamefont {S.}~\bibnamefont {Merkel}}, \bibinfo {author} {\bibfnamefont {T.}~\bibnamefont {Herr}}, \bibinfo {author} {\bibfnamefont {A.}~\bibnamefont {Silberfarb}}, \bibinfo {author} {\bibfnamefont {I.~H.}\ \bibnamefont {Deutsch}},\ and\ \bibinfo {author} {\bibfnamefont {P.~S.}\ \bibnamefont {Jessen}},\ }\bibfield  {title} {\bibinfo {title} {Quantum control of the hyperfine spin of a {Cs} atom ensemble},\ }\href@noop {} {\bibfield  {journal} {\bibinfo  {journal} {Physical Review Letters}\ }\textbf {\bibinfo {volume} {99}},\ \bibinfo {pages} {163002} (\bibinfo {year} {2007})}\BibitemShut {NoStop}%
\bibitem [{\citenamefont {Omanakuttan}\ \emph {et~al.}(2021)\citenamefont {Omanakuttan}, \citenamefont {Mitra}, \citenamefont {Martin},\ and\ \citenamefont {Deutsch}}]{Omanakuttan2021}%
  \BibitemOpen
  \bibfield  {author} {\bibinfo {author} {\bibfnamefont {S.}~\bibnamefont {Omanakuttan}}, \bibinfo {author} {\bibfnamefont {A.}~\bibnamefont {Mitra}}, \bibinfo {author} {\bibfnamefont {M.~J.}\ \bibnamefont {Martin}},\ and\ \bibinfo {author} {\bibfnamefont {I.~H.}\ \bibnamefont {Deutsch}},\ }\bibfield  {title} {\bibinfo {title} {Quantum optimal control of ten-level nuclear spin qudits in {$^{87}$Sr}},\ }\href@noop {} {\bibfield  {journal} {\bibinfo  {journal} {Physical Review A}\ }\textbf {\bibinfo {volume} {104}},\ \bibinfo {pages} {L060401} (\bibinfo {year} {2021})}\BibitemShut {NoStop}%
\bibitem [{\citenamefont {James}\ and\ \citenamefont {Jerke}(2007)}]{James2007}%
  \BibitemOpen
  \bibfield  {author} {\bibinfo {author} {\bibfnamefont {D.~F.}\ \bibnamefont {James}}\ and\ \bibinfo {author} {\bibfnamefont {J.}~\bibnamefont {Jerke}},\ }\bibfield  {title} {\bibinfo {title} {Effective {H}amiltonian theory and its applications in quantum information},\ }\href@noop {} {\bibfield  {journal} {\bibinfo  {journal} {Canadian Journal of Physics}\ }\textbf {\bibinfo {volume} {85}},\ \bibinfo {pages} {625} (\bibinfo {year} {2007})}\BibitemShut {NoStop}%
\bibitem [{\citenamefont {Steck}()}]{Steck}%
  \BibitemOpen
  \bibfield  {author} {\bibinfo {author} {\bibfnamefont {D.~A.}\ \bibnamefont {Steck}},\ }\href@noop {} {\bibinfo {title} {Quantum and atom optics}},\ \bibinfo {howpublished} {Available online at http://steck.us/teaching}\BibitemShut {NoStop}%
\bibitem [{\citenamefont {Deutsch}\ and\ \citenamefont {Jessen}(2010)}]{Deutsch2010}%
  \BibitemOpen
  \bibfield  {author} {\bibinfo {author} {\bibfnamefont {I.~H.}\ \bibnamefont {Deutsch}}\ and\ \bibinfo {author} {\bibfnamefont {P.~S.}\ \bibnamefont {Jessen}},\ }\bibfield  {title} {\bibinfo {title} {Quantum control and measurement of atomic spins in polarization spectroscopy},\ }\href@noop {} {\bibfield  {journal} {\bibinfo  {journal} {Optics Communications}\ }\textbf {\bibinfo {volume} {283}},\ \bibinfo {pages} {681} (\bibinfo {year} {2010})}\BibitemShut {NoStop}%
\bibitem [{\citenamefont {Le~Kien}\ \emph {et~al.}(2013)\citenamefont {Le~Kien}, \citenamefont {Schneeweiss},\ and\ \citenamefont {Rauschenbeutel}}]{LeKien2013}%
  \BibitemOpen
  \bibfield  {author} {\bibinfo {author} {\bibfnamefont {F.}~\bibnamefont {Le~Kien}}, \bibinfo {author} {\bibfnamefont {P.}~\bibnamefont {Schneeweiss}},\ and\ \bibinfo {author} {\bibfnamefont {A.}~\bibnamefont {Rauschenbeutel}},\ }\bibfield  {title} {\bibinfo {title} {Dynamical polarizability of atoms in arbitrary light fields: general theory and application to cesium},\ }\href@noop {} {\bibfield  {journal} {\bibinfo  {journal} {The European Physical Journal D}\ }\textbf {\bibinfo {volume} {67}},\ \bibinfo {pages} {92} (\bibinfo {year} {2013})}\BibitemShut {NoStop}%
\bibitem [{\citenamefont {Yang}\ \emph {et~al.}(2024)\citenamefont {Yang}, \citenamefont {Luo}, \citenamefont {Zhang}, \citenamefont {Wang}, \citenamefont {Zou}, \citenamefont {Xia},\ and\ \citenamefont {Lu}}]{Yang2024}%
  \BibitemOpen
  \bibfield  {author} {\bibinfo {author} {\bibfnamefont {Y.~A.}\ \bibnamefont {Yang}}, \bibinfo {author} {\bibfnamefont {W.-T.}\ \bibnamefont {Luo}}, \bibinfo {author} {\bibfnamefont {J.-L.}\ \bibnamefont {Zhang}}, \bibinfo {author} {\bibfnamefont {S.-Z.}\ \bibnamefont {Wang}}, \bibinfo {author} {\bibfnamefont {C.-L.}\ \bibnamefont {Zou}}, \bibinfo {author} {\bibfnamefont {T.}~\bibnamefont {Xia}},\ and\ \bibinfo {author} {\bibfnamefont {Z.-T.}\ \bibnamefont {Lu}},\ }\bibfield  {title} {\bibinfo {title} {Minute-scale {S}chr{\"o}dinger-cat state of spin-5/2 atoms},\ }\href@noop {} {\bibfield  {journal} {\bibinfo  {journal} {Nature Photonics}\ } (\bibinfo {year} {2024})}\BibitemShut {NoStop}%
\bibitem [{\citenamefont {Kusano}\ \emph {et~al.}(2026)\citenamefont {Kusano}, \citenamefont {Shibata}, \citenamefont {Yeh}, \citenamefont {Saito}, \citenamefont {Nakamura}, \citenamefont {Yokoyama}, \citenamefont {Kashimoto}, \citenamefont {Takano}, \citenamefont {Takasu}, \citenamefont {Takagi},\ and\ \citenamefont {Takahashi}}]{Kusano2026SpinCat}%
  \BibitemOpen
  \bibfield  {author} {\bibinfo {author} {\bibfnamefont {T.}~\bibnamefont {Kusano}}, \bibinfo {author} {\bibfnamefont {K.}~\bibnamefont {Shibata}}, \bibinfo {author} {\bibfnamefont {C.-H.}\ \bibnamefont {Yeh}}, \bibinfo {author} {\bibfnamefont {K.}~\bibnamefont {Saito}}, \bibinfo {author} {\bibfnamefont {Y.}~\bibnamefont {Nakamura}}, \bibinfo {author} {\bibfnamefont {R.}~\bibnamefont {Yokoyama}}, \bibinfo {author} {\bibfnamefont {T.}~\bibnamefont {Kashimoto}}, \bibinfo {author} {\bibfnamefont {T.}~\bibnamefont {Takano}}, \bibinfo {author} {\bibfnamefont {Y.}~\bibnamefont {Takasu}}, \bibinfo {author} {\bibfnamefont {R.}~\bibnamefont {Takagi}},\ and\ \bibinfo {author} {\bibfnamefont {Y.}~\bibnamefont {Takahashi}},\ }\href@noop {} {\bibinfo {title} {Spin-cat qubit with biased noise in an optical tweezer array}} (\bibinfo {year} {2026}),\ \Eprint {https://arxiv.org/abs/2602.22883} {arXiv:2602.22883 [quant-ph]} \BibitemShut {NoStop}%
\bibitem [{\citenamefont {Ozeri}\ \emph {et~al.}(2007)\citenamefont {Ozeri}, \citenamefont {Itano}, \citenamefont {Blakestad}, \citenamefont {Britton}, \citenamefont {Chiaverini}, \citenamefont {Jost}, \citenamefont {Langer}, \citenamefont {Leibfried}, \citenamefont {Reichle}, \citenamefont {Seidelin}, \citenamefont {Wesenberg},\ and\ \citenamefont {Wineland}}]{Ozeri2007}%
  \BibitemOpen
  \bibfield  {author} {\bibinfo {author} {\bibfnamefont {R.}~\bibnamefont {Ozeri}}, \bibinfo {author} {\bibfnamefont {W.~M.}\ \bibnamefont {Itano}}, \bibinfo {author} {\bibfnamefont {R.~B.}\ \bibnamefont {Blakestad}}, \bibinfo {author} {\bibfnamefont {J.}~\bibnamefont {Britton}}, \bibinfo {author} {\bibfnamefont {J.}~\bibnamefont {Chiaverini}}, \bibinfo {author} {\bibfnamefont {J.~D.}\ \bibnamefont {Jost}}, \bibinfo {author} {\bibfnamefont {C.}~\bibnamefont {Langer}}, \bibinfo {author} {\bibfnamefont {D.}~\bibnamefont {Leibfried}}, \bibinfo {author} {\bibfnamefont {R.}~\bibnamefont {Reichle}}, \bibinfo {author} {\bibfnamefont {S.}~\bibnamefont {Seidelin}}, \bibinfo {author} {\bibfnamefont {J.~H.}\ \bibnamefont {Wesenberg}},\ and\ \bibinfo {author} {\bibfnamefont {D.~J.}\ \bibnamefont {Wineland}},\ }\bibfield  {title} {\bibinfo {title} {Errors in trapped-ion quantum gates due to spontaneous photon scattering},\ }\href@noop {} {\bibfield  {journal} {\bibinfo  {journal} {Physical Review A}\ }\textbf {\bibinfo
  {volume} {75}},\ \bibinfo {pages} {042329} (\bibinfo {year} {2007})}\BibitemShut {NoStop}%
\bibitem [{\citenamefont {Moore}\ \emph {et~al.}(2023)\citenamefont {Moore}, \citenamefont {Campbell}, \citenamefont {Hudson}, \citenamefont {Boguslawski}, \citenamefont {Wineland},\ and\ \citenamefont {Allcock}}]{Moore2023}%
  \BibitemOpen
  \bibfield  {author} {\bibinfo {author} {\bibfnamefont {I.~D.}\ \bibnamefont {Moore}}, \bibinfo {author} {\bibfnamefont {W.~C.}\ \bibnamefont {Campbell}}, \bibinfo {author} {\bibfnamefont {E.~R.}\ \bibnamefont {Hudson}}, \bibinfo {author} {\bibfnamefont {M.~J.}\ \bibnamefont {Boguslawski}}, \bibinfo {author} {\bibfnamefont {D.~J.}\ \bibnamefont {Wineland}},\ and\ \bibinfo {author} {\bibfnamefont {D.~T.~C.}\ \bibnamefont {Allcock}},\ }\bibfield  {title} {\bibinfo {title} {Photon scattering errors during stimulated {R}aman transitions in trapped-ion qubits},\ }\href@noop {} {\bibfield  {journal} {\bibinfo  {journal} {Physical Review A}\ }\textbf {\bibinfo {volume} {107}},\ \bibinfo {pages} {032413} (\bibinfo {year} {2023})}\BibitemShut {NoStop}%
\bibitem [{\citenamefont {Low}\ \emph {et~al.}(2020)\citenamefont {Low}, \citenamefont {White}, \citenamefont {Cox}, \citenamefont {Day},\ and\ \citenamefont {Senko}}]{Low2020}%
  \BibitemOpen
  \bibfield  {author} {\bibinfo {author} {\bibfnamefont {P.~J.}\ \bibnamefont {Low}}, \bibinfo {author} {\bibfnamefont {B.~M.}\ \bibnamefont {White}}, \bibinfo {author} {\bibfnamefont {A.~A.}\ \bibnamefont {Cox}}, \bibinfo {author} {\bibfnamefont {M.~L.}\ \bibnamefont {Day}},\ and\ \bibinfo {author} {\bibfnamefont {C.}~\bibnamefont {Senko}},\ }\bibfield  {title} {\bibinfo {title} {Practical trapped-ion protocols for universal qudit-based quantum computing},\ }\href@noop {} {\bibfield  {journal} {\bibinfo  {journal} {Physical Review Research}\ }\textbf {\bibinfo {volume} {2}},\ \bibinfo {pages} {033128} (\bibinfo {year} {2020})}\BibitemShut {NoStop}%
\bibitem [{\citenamefont {Leibfried}\ \emph {et~al.}(2003)\citenamefont {Leibfried}, \citenamefont {Blatt}, \citenamefont {Monroe},\ and\ \citenamefont {Wineland}}]{Leibfried2003}%
  \BibitemOpen
  \bibfield  {author} {\bibinfo {author} {\bibfnamefont {D.}~\bibnamefont {Leibfried}}, \bibinfo {author} {\bibfnamefont {R.}~\bibnamefont {Blatt}}, \bibinfo {author} {\bibfnamefont {C.}~\bibnamefont {Monroe}},\ and\ \bibinfo {author} {\bibfnamefont {D.}~\bibnamefont {Wineland}},\ }\bibfield  {title} {\bibinfo {title} {Quantum dynamics of single trapped ions},\ }\href@noop {} {\bibfield  {journal} {\bibinfo  {journal} {Reviews of Modern Physics}\ }\textbf {\bibinfo {volume} {75}},\ \bibinfo {pages} {281} (\bibinfo {year} {2003})}\BibitemShut {NoStop}%
\bibitem [{\citenamefont {Omanakuttan}\ \emph {et~al.}(2023)\citenamefont {Omanakuttan}, \citenamefont {Mitra}, \citenamefont {Meier}, \citenamefont {Martin},\ and\ \citenamefont {Deutsch}}]{Omanakuttan2023}%
  \BibitemOpen
  \bibfield  {author} {\bibinfo {author} {\bibfnamefont {S.}~\bibnamefont {Omanakuttan}}, \bibinfo {author} {\bibfnamefont {A.}~\bibnamefont {Mitra}}, \bibinfo {author} {\bibfnamefont {E.~J.}\ \bibnamefont {Meier}}, \bibinfo {author} {\bibfnamefont {M.~J.}\ \bibnamefont {Martin}},\ and\ \bibinfo {author} {\bibfnamefont {I.~H.}\ \bibnamefont {Deutsch}},\ }\bibfield  {title} {\bibinfo {title} {Qudit entanglers using quantum optimal control},\ }\href@noop {} {\bibfield  {journal} {\bibinfo  {journal} {PRX Quantum}\ }\textbf {\bibinfo {volume} {4}},\ \bibinfo {pages} {040333} (\bibinfo {year} {2023})}\BibitemShut {NoStop}%
\bibitem [{\citenamefont {Onggadinata}\ \emph {et~al.}(2026)\citenamefont {Onggadinata}, \citenamefont {Koh}, \citenamefont {Maity}, \citenamefont {Goh}, \citenamefont {Weber}, \citenamefont {Lim}, \citenamefont {Ng},\ and\ \citenamefont {Koh}}]{Onggadinata2026}%
  \BibitemOpen
  \bibfield  {author} {\bibinfo {author} {\bibfnamefont {K.}~\bibnamefont {Onggadinata}}, \bibinfo {author} {\bibfnamefont {S.~Y.}\ \bibnamefont {Koh}}, \bibinfo {author} {\bibfnamefont {A.}~\bibnamefont {Maity}}, \bibinfo {author} {\bibfnamefont {K.~E.~J.}\ \bibnamefont {Goh}}, \bibinfo {author} {\bibfnamefont {B.}~\bibnamefont {Weber}}, \bibinfo {author} {\bibfnamefont {K.~J.}\ \bibnamefont {Lim}}, \bibinfo {author} {\bibfnamefont {H.~K.}\ \bibnamefont {Ng}},\ and\ \bibinfo {author} {\bibfnamefont {T.~S.}\ \bibnamefont {Koh}},\ }\href@noop {} {\bibinfo {title} {Towards fault-tolerance with universal phase-error-transparent gates for high-spin cat codes}} (\bibinfo {year} {2026}),\ \Eprint {https://arxiv.org/abs/2608.05992} {arXiv:2608.05992 [quant-ph]} \BibitemShut {NoStop}%
\bibitem [{\citenamefont {Qichen}\ \emph {et~al.}(2026)\citenamefont {Qichen}, \citenamefont {Zhao}, \citenamefont {Lee}, \citenamefont {Zhang}, \citenamefont {Jayjong}, \citenamefont {Arnold},\ and\ \citenamefont {Barrett}}]{Qichen2026}%
  \BibitemOpen
  \bibfield  {author} {\bibinfo {author} {\bibfnamefont {Q.}~\bibnamefont {Qichen}}, \bibinfo {author} {\bibfnamefont {Q.}~\bibnamefont {Zhao}}, \bibinfo {author} {\bibfnamefont {M.~D.~K.}\ \bibnamefont {Lee}}, \bibinfo {author} {\bibfnamefont {Z.}~\bibnamefont {Zhang}}, \bibinfo {author} {\bibfnamefont {N.}~\bibnamefont {Jayjong}}, \bibinfo {author} {\bibfnamefont {K.~J.}\ \bibnamefont {Arnold}},\ and\ \bibinfo {author} {\bibfnamefont {M.~D.}\ \bibnamefont {Barrett}},\ }\bibfield  {title} {\bibinfo {title} {Zeeman degenerate sideband cooling in {$^{176}$Lu$^{+}$}},\ }\href@noop {} {\bibfield  {journal} {\bibinfo  {journal} {Physical Review A}\ }\textbf {\bibinfo {volume} {113}},\ \bibinfo {pages} {013109} (\bibinfo {year} {2026})}\BibitemShut {NoStop}%
\bibitem [{\citenamefont {Benhelm}\ \emph {et~al.}(2007)\citenamefont {Benhelm}, \citenamefont {Kirchmair}, \citenamefont {Rapol}, \citenamefont {Koerber}, \citenamefont {Roos},\ and\ \citenamefont {Blatt}}]{Benhelm2007}%
  \BibitemOpen
  \bibfield  {author} {\bibinfo {author} {\bibfnamefont {J.}~\bibnamefont {Benhelm}}, \bibinfo {author} {\bibfnamefont {G.}~\bibnamefont {Kirchmair}}, \bibinfo {author} {\bibfnamefont {U.}~\bibnamefont {Rapol}}, \bibinfo {author} {\bibfnamefont {T.}~\bibnamefont {Koerber}}, \bibinfo {author} {\bibfnamefont {C.~F.}\ \bibnamefont {Roos}},\ and\ \bibinfo {author} {\bibfnamefont {R.}~\bibnamefont {Blatt}},\ }\bibfield  {title} {\bibinfo {title} {Measurement of the hyperfine structure of the {$S_{1/2}$--$D_{5/2}$} transition in {$^{43}$Ca$^{+}$}},\ }\href@noop {} {\bibfield  {journal} {\bibinfo  {journal} {Physical Review A}\ }\textbf {\bibinfo {volume} {75}},\ \bibinfo {pages} {032506} (\bibinfo {year} {2007})}\BibitemShut {NoStop}%
\bibitem [{\citenamefont {Tan}\ \emph {et~al.}(2015)\citenamefont {Tan}, \citenamefont {Gaebler}, \citenamefont {Lin}, \citenamefont {Wan}, \citenamefont {Bowler}, \citenamefont {Leibfried},\ and\ \citenamefont {Wineland}}]{Tan2015}%
  \BibitemOpen
  \bibfield  {author} {\bibinfo {author} {\bibfnamefont {T.~R.}\ \bibnamefont {Tan}}, \bibinfo {author} {\bibfnamefont {J.~P.}\ \bibnamefont {Gaebler}}, \bibinfo {author} {\bibfnamefont {Y.}~\bibnamefont {Lin}}, \bibinfo {author} {\bibfnamefont {Y.}~\bibnamefont {Wan}}, \bibinfo {author} {\bibfnamefont {R.}~\bibnamefont {Bowler}}, \bibinfo {author} {\bibfnamefont {D.}~\bibnamefont {Leibfried}},\ and\ \bibinfo {author} {\bibfnamefont {D.~J.}\ \bibnamefont {Wineland}},\ }\bibfield  {title} {\bibinfo {title} {Multi-element logic gates for trapped-ion qubits},\ }\href@noop {} {\bibfield  {journal} {\bibinfo  {journal} {Nature}\ }\textbf {\bibinfo {volume} {528}},\ \bibinfo {pages} {380} (\bibinfo {year} {2015})}\BibitemShut {NoStop}%
\bibitem [{\citenamefont {Heris}\ \emph {et~al.}(2026)\citenamefont {Heris}, \citenamefont {Liu},\ and\ \citenamefont {Mueller}}]{Heris2026}%
  \BibitemOpen
  \bibfield  {author} {\bibinfo {author} {\bibfnamefont {M.~H.}\ \bibnamefont {Heris}}, \bibinfo {author} {\bibfnamefont {Y.}~\bibnamefont {Liu}},\ and\ \bibinfo {author} {\bibfnamefont {F.}~\bibnamefont {Mueller}},\ }\href@noop {} {\bibinfo {title} {A hardware-native trapped-ion hybrid qudit-oscillator instruction set}} (\bibinfo {year} {2026}),\ \Eprint {https://arxiv.org/abs/2610.13843} {arXiv:2610.13843 [quant-ph]} \BibitemShut {NoStop}%
\bibitem [{\citenamefont {Lloyd}(1995)}]{Lloyd1995}%
  \BibitemOpen
  \bibfield  {author} {\bibinfo {author} {\bibfnamefont {S.}~\bibnamefont {Lloyd}},\ }\bibfield  {title} {\bibinfo {title} {Almost any quantum logic gate is universal},\ }\href@noop {} {\bibfield  {journal} {\bibinfo  {journal} {Physical Review Letters}\ }\textbf {\bibinfo {volume} {75}},\ \bibinfo {pages} {346} (\bibinfo {year} {1995})}\BibitemShut {NoStop}%
\bibitem [{\citenamefont {Ball}\ \emph {et~al.}(2021)\citenamefont {Ball}, \citenamefont {Biercuk}, \citenamefont {Carvalho}, \citenamefont {Chen}, \citenamefont {Hush}, \citenamefont {De~Castro}, \citenamefont {Li}, \citenamefont {Liebermann}, \citenamefont {Slatyer}, \citenamefont {Edmunds}, \citenamefont {Frey}, \citenamefont {Hempel},\ and\ \citenamefont {Milne}}]{Ball2021}%
  \BibitemOpen
  \bibfield  {author} {\bibinfo {author} {\bibfnamefont {H.}~\bibnamefont {Ball}}, \bibinfo {author} {\bibfnamefont {M.~J.}\ \bibnamefont {Biercuk}}, \bibinfo {author} {\bibfnamefont {A.~R.~R.}\ \bibnamefont {Carvalho}}, \bibinfo {author} {\bibfnamefont {J.}~\bibnamefont {Chen}}, \bibinfo {author} {\bibfnamefont {M.}~\bibnamefont {Hush}}, \bibinfo {author} {\bibfnamefont {L.~A.}\ \bibnamefont {De~Castro}}, \bibinfo {author} {\bibfnamefont {L.}~\bibnamefont {Li}}, \bibinfo {author} {\bibfnamefont {P.~J.}\ \bibnamefont {Liebermann}}, \bibinfo {author} {\bibfnamefont {H.~J.}\ \bibnamefont {Slatyer}}, \bibinfo {author} {\bibfnamefont {C.}~\bibnamefont {Edmunds}}, \bibinfo {author} {\bibfnamefont {V.}~\bibnamefont {Frey}}, \bibinfo {author} {\bibfnamefont {C.}~\bibnamefont {Hempel}},\ and\ \bibinfo {author} {\bibfnamefont {A.}~\bibnamefont {Milne}},\ }\bibfield  {title} {\bibinfo {title} {Software tools for quantum control: improving quantum computer performance through noise and error suppression},\ }\href@noop
  {} {\bibfield  {journal} {\bibinfo  {journal} {Quantum Science and Technology}\ }\textbf {\bibinfo {volume} {6}},\ \bibinfo {pages} {044011} (\bibinfo {year} {2021})}\BibitemShut {NoStop}%
\bibitem [{\citenamefont {Bentley}\ \emph {et~al.}(2020)\citenamefont {Bentley}, \citenamefont {Ball}, \citenamefont {Biercuk}, \citenamefont {Carvalho}, \citenamefont {Hush},\ and\ \citenamefont {Slatyer}}]{Bentley2020}%
  \BibitemOpen
  \bibfield  {author} {\bibinfo {author} {\bibfnamefont {C.~D.~B.}\ \bibnamefont {Bentley}}, \bibinfo {author} {\bibfnamefont {H.}~\bibnamefont {Ball}}, \bibinfo {author} {\bibfnamefont {M.~J.}\ \bibnamefont {Biercuk}}, \bibinfo {author} {\bibfnamefont {A.~R.~R.}\ \bibnamefont {Carvalho}}, \bibinfo {author} {\bibfnamefont {M.~R.}\ \bibnamefont {Hush}},\ and\ \bibinfo {author} {\bibfnamefont {H.~J.}\ \bibnamefont {Slatyer}},\ }\bibfield  {title} {\bibinfo {title} {Numeric optimization for configurable, parallel, error-robust entangling gates in large ion registers},\ }\href@noop {} {\bibfield  {journal} {\bibinfo  {journal} {Advanced Quantum Technologies}\ }\textbf {\bibinfo {volume} {3}},\ \bibinfo {pages} {2000044} (\bibinfo {year} {2020})}\BibitemShut {NoStop}%
\bibitem [{\citenamefont {Matsos}\ \emph {et~al.}(2024)\citenamefont {Matsos}, \citenamefont {Valahu}, \citenamefont {Navickas}, \citenamefont {Rao}, \citenamefont {Millican}, \citenamefont {Kolesnikow}, \citenamefont {Biercuk},\ and\ \citenamefont {Tan}}]{Matsos2024}%
  \BibitemOpen
  \bibfield  {author} {\bibinfo {author} {\bibfnamefont {V.}~\bibnamefont {Matsos}}, \bibinfo {author} {\bibfnamefont {C.}~\bibnamefont {Valahu}}, \bibinfo {author} {\bibfnamefont {T.}~\bibnamefont {Navickas}}, \bibinfo {author} {\bibfnamefont {A.}~\bibnamefont {Rao}}, \bibinfo {author} {\bibfnamefont {M.}~\bibnamefont {Millican}}, \bibinfo {author} {\bibfnamefont {X.}~\bibnamefont {Kolesnikow}}, \bibinfo {author} {\bibfnamefont {M.}~\bibnamefont {Biercuk}},\ and\ \bibinfo {author} {\bibfnamefont {T.}~\bibnamefont {Tan}},\ }\bibfield  {title} {\bibinfo {title} {Robust and deterministic preparation of bosonic logical states in a trapped ion},\ }\href@noop {} {\bibfield  {journal} {\bibinfo  {journal} {Physical Review Letters}\ }\textbf {\bibinfo {volume} {133}},\ \bibinfo {pages} {050602} (\bibinfo {year} {2024})}\BibitemShut {NoStop}%
\bibitem [{\citenamefont {Matsos}\ \emph {et~al.}(2025)\citenamefont {Matsos}, \citenamefont {Valahu}, \citenamefont {Millican}, \citenamefont {Navickas}, \citenamefont {Kolesnikow}, \citenamefont {Biercuk},\ and\ \citenamefont {Tan}}]{Matsos2025}%
  \BibitemOpen
  \bibfield  {author} {\bibinfo {author} {\bibfnamefont {V.~G.}\ \bibnamefont {Matsos}}, \bibinfo {author} {\bibfnamefont {C.~H.}\ \bibnamefont {Valahu}}, \bibinfo {author} {\bibfnamefont {M.~J.}\ \bibnamefont {Millican}}, \bibinfo {author} {\bibfnamefont {T.}~\bibnamefont {Navickas}}, \bibinfo {author} {\bibfnamefont {X.~C.}\ \bibnamefont {Kolesnikow}}, \bibinfo {author} {\bibfnamefont {M.~J.}\ \bibnamefont {Biercuk}},\ and\ \bibinfo {author} {\bibfnamefont {T.~R.}\ \bibnamefont {Tan}},\ }\bibfield  {title} {\bibinfo {title} {Universal quantum gate set for {G}ottesman--{K}itaev--{P}reskill logical qubits},\ }\href@noop {} {\bibfield  {journal} {\bibinfo  {journal} {Nature Physics}\ } (\bibinfo {year} {2025})}\BibitemShut {NoStop}%
\bibitem [{\citenamefont {Millican}\ \emph {et~al.}(2025)\citenamefont {Millican}, \citenamefont {Matsos}, \citenamefont {Valahu}, \citenamefont {Navickas}, \citenamefont {Bond},\ and\ \citenamefont {Tan}}]{Millican2025}%
  \BibitemOpen
  \bibfield  {author} {\bibinfo {author} {\bibfnamefont {M.~J.}\ \bibnamefont {Millican}}, \bibinfo {author} {\bibfnamefont {V.~G.}\ \bibnamefont {Matsos}}, \bibinfo {author} {\bibfnamefont {C.~H.}\ \bibnamefont {Valahu}}, \bibinfo {author} {\bibfnamefont {T.}~\bibnamefont {Navickas}}, \bibinfo {author} {\bibfnamefont {L.~J.}\ \bibnamefont {Bond}},\ and\ \bibinfo {author} {\bibfnamefont {T.~R.}\ \bibnamefont {Tan}},\ }\bibfield  {title} {\bibinfo {title} {Engineering continuous-variable entanglement in mechanical oscillators with optimal control},\ }\href@noop {} {\bibfield  {journal} {\bibinfo  {journal} {Physical Review Letters}\ }\textbf {\bibinfo {volume} {135}},\ \bibinfo {pages} {233604} (\bibinfo {year} {2025})}\BibitemShut {NoStop}%
\bibitem [{\citenamefont {Ozeri}\ \emph {et~al.}(2005)\citenamefont {Ozeri}, \citenamefont {Langer}, \citenamefont {Jost}, \citenamefont {DeMarco}, \citenamefont {Ben-Kish}, \citenamefont {Blakestad}, \citenamefont {Britton}, \citenamefont {Chiaverini}, \citenamefont {Itano}, \citenamefont {Hume}, \citenamefont {Leibfried}, \citenamefont {Rosenband}, \citenamefont {Schmidt},\ and\ \citenamefont {Wineland}}]{Ozeri2005}%
  \BibitemOpen
  \bibfield  {author} {\bibinfo {author} {\bibfnamefont {R.}~\bibnamefont {Ozeri}}, \bibinfo {author} {\bibfnamefont {C.}~\bibnamefont {Langer}}, \bibinfo {author} {\bibfnamefont {J.~D.}\ \bibnamefont {Jost}}, \bibinfo {author} {\bibfnamefont {B.}~\bibnamefont {DeMarco}}, \bibinfo {author} {\bibfnamefont {A.}~\bibnamefont {Ben-Kish}}, \bibinfo {author} {\bibfnamefont {B.~R.}\ \bibnamefont {Blakestad}}, \bibinfo {author} {\bibfnamefont {J.}~\bibnamefont {Britton}}, \bibinfo {author} {\bibfnamefont {J.}~\bibnamefont {Chiaverini}}, \bibinfo {author} {\bibfnamefont {W.~M.}\ \bibnamefont {Itano}}, \bibinfo {author} {\bibfnamefont {D.~B.}\ \bibnamefont {Hume}}, \bibinfo {author} {\bibfnamefont {D.}~\bibnamefont {Leibfried}}, \bibinfo {author} {\bibfnamefont {T.}~\bibnamefont {Rosenband}}, \bibinfo {author} {\bibfnamefont {P.~O.}\ \bibnamefont {Schmidt}},\ and\ \bibinfo {author} {\bibfnamefont {D.~J.}\ \bibnamefont {Wineland}},\ }\bibfield  {title} {\bibinfo {title} {Hyperfine coherence in the presence of
  spontaneous photon scattering},\ }\href@noop {} {\bibfield  {journal} {\bibinfo  {journal} {Physical Review Letters}\ }\textbf {\bibinfo {volume} {95}},\ \bibinfo {pages} {030403} (\bibinfo {year} {2005})}\BibitemShut {NoStop}%
\bibitem [{\citenamefont {Wineland}\ \emph {et~al.}(2003)\citenamefont {Wineland}, \citenamefont {Barrett}, \citenamefont {Britton}, \citenamefont {Chiaverini}, \citenamefont {DeMarco}, \citenamefont {Itano}, \citenamefont {Jelenkovi{\'c}}, \citenamefont {Langer}, \citenamefont {Leibfried}, \citenamefont {Meyer}, \citenamefont {Rosenband},\ and\ \citenamefont {Sch{\"a}tz}}]{Wineland2003}%
  \BibitemOpen
  \bibfield  {author} {\bibinfo {author} {\bibfnamefont {D.~J.}\ \bibnamefont {Wineland}}, \bibinfo {author} {\bibfnamefont {M.}~\bibnamefont {Barrett}}, \bibinfo {author} {\bibfnamefont {J.}~\bibnamefont {Britton}}, \bibinfo {author} {\bibfnamefont {J.}~\bibnamefont {Chiaverini}}, \bibinfo {author} {\bibfnamefont {B.}~\bibnamefont {DeMarco}}, \bibinfo {author} {\bibfnamefont {W.~M.}\ \bibnamefont {Itano}}, \bibinfo {author} {\bibfnamefont {B.}~\bibnamefont {Jelenkovi{\'c}}}, \bibinfo {author} {\bibfnamefont {C.}~\bibnamefont {Langer}}, \bibinfo {author} {\bibfnamefont {D.}~\bibnamefont {Leibfried}}, \bibinfo {author} {\bibfnamefont {V.}~\bibnamefont {Meyer}}, \bibinfo {author} {\bibfnamefont {T.}~\bibnamefont {Rosenband}},\ and\ \bibinfo {author} {\bibfnamefont {T.}~\bibnamefont {Sch{\"a}tz}},\ }\bibfield  {title} {\bibinfo {title} {Quantum information processing with trapped ions},\ }\href@noop {} {\bibfield  {journal} {\bibinfo  {journal} {Philosophical Transactions of the Royal Society of London. Series
  A: Mathematical, Physical and Engineering Sciences}\ }\textbf {\bibinfo {volume} {361}},\ \bibinfo {pages} {1349} (\bibinfo {year} {2003})}\BibitemShut {NoStop}%
\end{thebibliography}%

\end{document}